\documentclass[fleqn,usenatbib]{mnras}

\usepackage{newtxtext,newtxmath}
\usepackage{xfrac, nicefrac}

\usepackage[T1]{fontenc}

\DeclareRobustCommand{\VAN}[3]{#2}
\let\VANthebibliography\thebibliography
\def\thebibliography{\DeclareRobustCommand{\VAN}[3]{##3}\VANthebibliography}

\usepackage{graphicx}	
\usepackage{amsmath}	
\usepackage{caption}
\usepackage{subcaption}

\title[Dust in External Photoevaporative Winds]{A Particle-based Approach to Dust Dynamics in External Photoevaporative Winds}
\author[S. Paine et al.]{
S. Paine$^1$,
T. J. Haworth $^1$
and R. P. Nelson$^1$
\\
$^{1}$Astronomy Unit, School of Physics and Astronomy, Queen Mary University of London, London E1 4NS, UK\\
}

\date{Accepted XXX. Received YYY; in original form ZZZ}

\pubyear{2025}

\begin{document}
\label{firstpage}
\pagerange{\pageref{firstpage}--\pageref{lastpage}}
\maketitle

\begin{abstract}
Planet-forming discs in sufficiently strong UV environments lose gas in external photoevaporative winds. Dust can also be entrained within these winds, which has consequences for the possible solids reservoir for planet formation, and determines the shielding of the disc by the wind. This has previously been studied in 1D models, with predictions for the maximum entrained size, as well as a predicted population of stalled dust of decreasing grain size with distance from the disc. We wrote and tested a new dust particle solver to make the first study of the entrainment and dynamics of dust, using steady state solutions of state-of-the-art 1D and 2D radiation hydrodynamic simulations of externally photoevaporating discs. In our 1D models, we only consider the outer disc at the midplane, verifying previous studies. In our 2D simulations, the wind is launched from the disc surface, as well as the disc edge. In 2D we find that the maximum entrained grain size varies substantially with angle relative to the plane of the disc, from $\sim100$\,$\mu$m near the disc outer edge down to $\sim1$\,$\mu$m or even sub-micron in the weaker wind from the disc surface. The gradient of stalled dust seen in 1D also only appears near the disc outer edge in 2D, but not from the disc surface. This agrees qualitatively with observations of silhouette discs in the Orion Nebula Cluster. Despite the spatial variation of the dust, the extinction of the UV radiation remains fairly uniform due to the opacity being dominated by the small grains, and depends more on the dust distribution within the disc itself. 
\end{abstract}




\section{Introduction}

The star formation process results in a circumstellar, planet-forming disc of gas and dust around the young star.  The evolution of the dust is of particular interest, since it is the growth of solids that directly leads to terrestrial planet formation and is likely an important component of giant planet formation \citep[e.g.][]{2014prpl.conf..339T}.

Dust is expected to evolve quite differently from the gas, and in a manner that is highly dependent on the grain size (specifically, dependent upon the Stokes number). Whereas small grains closely follow the gas motions, gas drag can decouple the larger dust motions from that of the gas. In particular, pressure support in the disc means that the orbital velocity is slightly sub-Keplerian for the gas, but Keplerian for the dust. That results in a headwind that leads to radial drift of the dust inwards into the disc \citep[e.g.][]{1977MNRAS.180...57W}. This flux of larger dust grains can lead to planet formation by ``pebble accretion'' \citep{2012A&A...544A..32L, 2014A&A...572A..35L, 2017AREPS..45..359J}, but conversely if left unchecked can also deplete the dust in the disc on very short timescales. Substructures in the disc that result in pressure bumps, and fragmentation, may provide a means of retaining the dust for longer \citep[e.g.][]{2012A&A...539A.148B, 2018ApJ...869L..46D, 2021ApJ...921...72M}. 

Observations of discs with facilities such as ALMA and JWST are now able to directly detect this difference between the gas and dust distributions in discs and to demonstrate radial drift in action \citep[e.g.][]{2018ApJ...869L..41A, 2021ApJS..257....1O, 2023ApJ...957L..22B, 2025arXiv250104587G}. 

Some of the most important factors in understanding planet formation are the timescales on which planet-forming discs evolve, the key processes driving that evolution and where mass is distributed over time. It is therefore proving very important to understand whether angular momentum is redistributed by viscous transport or magnetohydrodynamic winds \citep{2022MNRAS.512L..74T, 2022MNRAS.512.2290T, 2023ASPC..534..539M, 2024MNRAS.527.7588C}. 

The role of photoevaporative winds is also thought to be a significant factor in disc evolution and lifetimes. Internal photoevaporation depletes the inner disc of material due to irradiation by the central star that the disc orbits \citep[for a review see][]{2023ASPC..534..567P} and external photoevaporation depletes the outer disc due to FUV irradiation by other stars in the star-forming cluster \citep[for a review, see][]{2022EPJP..137.1132W}. To first order, photoevaporative winds result when the gas is heated sufficiently that the local sound speed exceeds the escape velocity and that has been the subject of multiple analytic and numerical modelling efforts both for internal and external photoevaporation \citep[e.g.][]{1994ApJ...428..654H, 1998ApJ...499..758J, 2000ApJ...539..258R, 2004ApJ...611..360A, 2009ApJ...690.1539G, 2012MNRAS.422.1880O, 2018MNRAS.481..452H, 2018ApJ...857...57N, 2019MNRAS.487..691P, 2019ApJ...874...90W}. External photoevaporation was originally identified from strongly irradiated discs in Orion almost 30 years ago, which show a cometary proplyd morphology \citep[e.g.][]{1993ApJ...410..696O, 1994ApJ...436..194O, 2000AJ....119.2919B, 2008AJ....136.2136R, 2023ApJ...954..127B, 2024arXiv240312604A}, and has recently returned to attention and is increasingly thought to play an important role even down to quite weak external UV fields \citep[e.g.][]{2016ApJ...826L..15K, 2017MNRAS.468L.108H,  2023A&A...673L...2V}. 

Early work considering the effect of externally driven UV winds on dust concentrations and planet formation in the disc and wind was done by \cite{2005Throop&Bally}, who focused on the impact on the dust-to-gas ratio in the disc.
More recently, attention has turned back to the entrainment of dust in these photoevaporative winds. \cite{2016MNRAS.461..742H, 2016MNRAS.463.2725H} studied the entrainment of dust in a general planar non-rotating wind and then extended that to study the delivery and entrainment of dust \citep{2016MNRAS.463.2725H, 2021MNRAS.501.1127H} in the self-similar semi-analytic 2D-axisymmetric internal photoevaporative wind model of \cite{2016MNRAS.460.3044C}. They find that only small dust is entrained and larger dust either does not get delivered into the wind in the first place, or if it were, would be dragged out a small way before falling back down onto the disc.

The entrainment of dust in external photoevaporative winds has been subject to less study. The first estimate of entrained dust sizes for T Tauri stars was studied in \cite{2005Takeuchi+}. Later, \cite{2016MNRAS.457.3593F} constructed 1D semi-analytic models of external photoevaporative winds and solved for the maximum entrained grain size, defined as the largest size that would reach force balance beyond a critical point in the flow. Similar to internal winds, they found that only small dust grains are expected to be entrained in the wind. This has a series of very important consequences for external photoevaporation. Firstly, if grain growth and radial drift has occurred in the outer disc then it will deplete the small grain abundance there, which means that only a small amount of dust is dragged out with the wind compared to if the dust distribution were like that in the interstellar medium (ISM). In the 1D models this effectively lowers the extinction in the wind leading to higher mass loss rates \citep{2016MNRAS.457.3593F, 2023MNRAS.526.4315H}. Secondly the entrainment of dust is also important because it reduces the solids mass reservoir for planet formation \citep[e.g.][]{2022A&A...666A..73B, 2023MNRAS.522.1939Q}. Finally, dust entrainment is an important ingredient for understanding the global evolution of discs based on observations of their dust content \citep{2020MNRAS.492.1279S}.

The conclusion that small grains are entrained was also recently supported by \cite{2021MNRAS.508.2493O} who ran isothermal slim-disc models of external photoevaporation, including a dust model. \cite{2023arXiv231020214G} and \cite{2023A&A...679A..15G} also recently demonstrated that substructures within discs can significantly affect the evolution of photoevaporating discs and coupled with fragmentation in dust over-densities can also replenish a small grain population to entrain into the wind. 

Observationally there is currently only a small amount of evidence available to demonstrate the entrainment of dust in external photoevaporative winds. \cite{2012ApJ...757...78M} studied the extended disc 114-426 in the Orion Nebular Cluster (ONC) finding a gradient in the maximum grain size in the outer disc, which is what would be expected from entrainment in an external photoevaporative wind. \cite{2001ApJ...561..830G} also compared hydrodynamical models with 11.7\,$\mu$m mid-infrared observations from \cite{1994ApJ...433..157H} suggesting that the dust opacity was lower than in the typical ISM by a factor $\sim30$. JWST should provide further multi-wavelength constraints for similar studies of grain entrainment. 

Overall, entrainment (or lack of it) in external photoevaporative winds is expected to be very important for discs with high mass loss rates. To date though, the theory of grain entrainment in external photoevaporative winds has been limited to 1D calculations that consider the flow from the disc outer midplane only. In reality around half of the mass loss is expected from the disc surface \citep{2019MNRAS.485.3895H}. If the entrained dust were to not follow the gas, for example falling down onto the midplane, then the extinction along different lines of sight to the disc may differ substantially from what is currently assumed. It is also very difficult to compare with observations using 1D models. In this paper we therefore develop a particle dynamics solver and apply it to 1D-radial and 2D-axisymmetric simulations of externally photoevaporating discs. Our goals in this paper are to test our existing understanding of grain entrainment in external photoevaporative winds, and see how things differ away from the midplane considered by the 1D models. 

This paper is organised as follows. In section 2 we introduce our particle solver and demonstrate its veracity using a series of benchmark tests. In section 3 we test the 1D semi-analytic models of \cite{2016MNRAS.457.3593F} and present the first applications to 2D-axisymmetric external photoevaporation calculations. In section 4 we discuss the implications of our results and a summary and conclusion is given in section 5.




\section{Numerical method}

\subsection{Overview}

In this paper, we add dust after the fact to pre-computed steady state radiation hydrodynamic simulations of discs undergoing external photoevaporation calculated with the \textsc{torus-3dpdr} code \citep[for details on the code, and previous applications to external photoevaporation see e.g.][]{2015MNRAS.454.2828B, 2019A&C....27...63H, 2019MNRAS.485.3895H,  2023MNRAS.526.4315H}. The details of the radiation hydrodynamic models are given at relevant points in section \ref{sec:results}, but generally consist of an imposed disc boundary condition and a dynamically evolved wind. We include models of 1D-radial and 2D-axisymmetric scenarios. We provide a brief overview of the methodology for solving the dust trajectories here and include further details in appendix \ref{appendix:method}. 

We developed a new particle solver code from scratch to evolve dust trajectories in these steady-state solutions of external photoevaporative winds. It uses a 2nd-order leapfrog-like integrator. Dust of a given size is introduced near the base of the wind and the equation of motion is solved, accounting for the gravitational, drag and centrifugal forces on the dust particle as it travels through the medium. These dust grains are kept at their starting size throughout the simulation, which we justify on the basis that the wind density is much lower than the disc and the dust-to-gas ratio in the wind is also expected to be depleted \citep{2001ApJ...561..830G, 2012ApJ...757...78M, 2016MNRAS.457.3593F} and so significant grain growth is not expected on a flow timescale. We will study the structure and possible fragmentation of the grains themselves in the wind in subsequent work, but from a dynamics perspective if the dust were to fragment into smaller particles they would simply be expected to follow the gas streamlines more closely. \cite{2016MNRAS.461..742H} found the back-reaction of dust entrained in internal photoevaporative winds (driven by the disc host star) to have no effect on the gas velocity and only cause a 5\% decrease of the gas density. We therefore do not consider any back-reaction of the dust on the gas in this paper.

We benchmarked the code in three parts: gravity, drag force and both combined; to ensure all its components all worked properly separately as well as together. These are briefly summarised below, and the methodology discussed in full detail in appendix \ref{appendix:method}.

\subsection{Benchmarks}

\subsubsection{Gravity}
We begin by benchmarking the gravity on its own. Defining $\hat{R}$, $\hat{\phi}$ and $\hat{z}$ as the radial, angular and vertical positions, respectively.
We then derive the equations of motion for gravity 
\begin{align}
    \left\langle \ddot{R}, R\ddot{\phi}, \ddot{z} \right\rangle &= \left\langle -\frac{GM_*R}{(R^2 + z^2)^{3/2}} + \frac{h^2}{R^3}, -2\dot{R}\dot{\phi},  -\frac{GM_*z}{(R^2 + z^2)^{3/2}} \right\rangle
\end{align}
where $M_*$ is the mass of the central star, $G$ is the gravitational constant, $h$ is the angular momentum of the particle and the dot \ $\dot{}$ \ superscripts represent derivatives with respect to time.
To benchmark, we placed particles on various Keplerian orbits and checked that they were stable and their energies and angular momenta were conserved, varying by less than 0.25\% over 10\,Myr timescales. 

\subsubsection{Aerodynamic Drag}
The next force to benchmark was aerodynamic drag. Here, we used the seminal paper of \cite{1977MNRAS.180...57W} as a benchmark to compare against. This places dust in a medium with a gas pressure gradient, introducing an aerodynamic force that opposes particle motion, relative to the gas it is in. This force equalises the dust's velocity with that of the local gas, but since the gas's orbital velocity is sub-Keplerian, the dust moves to a smaller orbit to compensate - this radial drift inwards causes grains to eventually be captured by their host stars (assuming no pressure bumps or other phenomena that would trap the dust).

For the remainder of this paper, we will only consider the Epstein drag regime, although the Stokes regime receives treatment in the appendix. The Epstein drag force and acceleration are
\begin{align}
    \vec{F}_D &= \frac{4\pi}{3} \rho s^2 \bar{v}\Delta\vec{v} \label{eqn:epstein}\\
    \vec{a}_D &= \frac{\rho}{s\rho_s} \bar{v} \Delta\vec{v}
\end{align}
where the grain radius and density are denoted by $s$ and $\rho_s$, respectively, while the gas density is represented without a subscript $\rho$.
$\bar{v}$ is the mean thermal velocity of the gas and $\Delta\vec{v}$ is the vector velocity difference between the gas and dust velocities.
Throughout this paper, we assume spherical dust grains with uniform density, to get a dust mass of $m_d = \nicefrac{4\pi}{3} \rho_s s^3$. Throughout this paper, we will also use grain `radius' and `size' interchangeably, with both referring to the radius of the dust grain.
Adding this force to our equations of motion gives us the final set of equations we will be solving to simulate dust dynamics:
\begin{align}
    \ddot{R} &= -\frac{GM_*R}{(R^2 + z^2)^{3/2}} + \frac{h^2}{R^3} + \frac{\rho}{s\rho_s} \bar{v}\Delta v_R\\
    R\ddot{\phi} &= \frac{\rho}{s\rho_s} \bar{v}\Delta v_\phi - 2\dot{R}\frac{h}{R^2}\\
    \ddot{z} &= -\frac{GM_*z}{(R^2 + z^2)^{3/2}} + \frac{\rho}{s\rho_s} \bar{v}\Delta v_z.
\end{align}

We then used the same disc model parameters used by the \cite{1977MNRAS.180...57W} paper, with power law distributions for the surface density
\begin{equation}
    \Sigma(R) = \Sigma_0 \left(R/R_0\right)^{-a+1}
\end{equation}
and the temperature 
\begin{equation}
    T(R) = T_0 \left(R/R_0\right)^{-m}.
\label{eqn:temperature}
\end{equation}
We select $M_*=1M_\odot$, $m=1$, $a=2$, $R_0=1\mathrm{AU}$, $T_0=600$K and $\Sigma_0 = 1000 \mathrm{g/cm^2}$ for our initial parameters.
The gas in these discs is supported by a pressure gradient, making the gas orbit at sub-Keplerian velocities. This velocity difference is responsible for the drag force on the dust, causing the grains to spiral in at different speeds based on their starting distances and sizes. Generally, the smallest dust sizes remain in the Epstein regime, quickly matching the local gas azimuthal velocity and reaching a terminal radial velocity.
\begin{figure}
    \centering
    \includegraphics[width=\columnwidth]{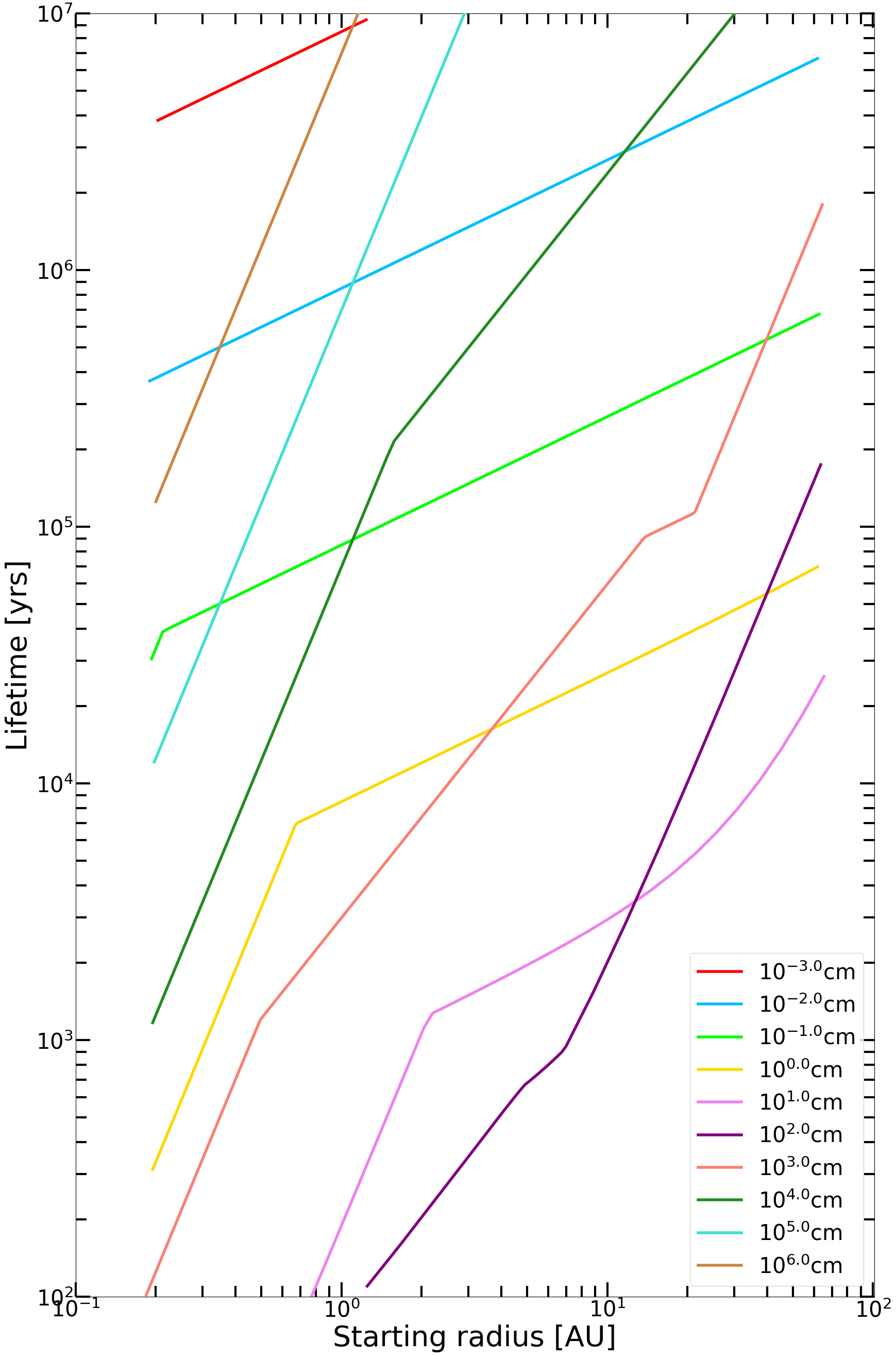} 
    \caption{Lifetime of various dust sizes, from different starting positions in a \protect\cite{1977MNRAS.180...57W} style disc. Here, the lifetime is calculated as the reciprocal of the terminal radial velocity of the dust: $\tau = \left(\frac{dR}{dt}\right)^{-1}$. On average, our numerical models agree with those in \protect\cite{1977MNRAS.180...57W} to within 3\%. The following parameters were used to reproduce the plot: $M_*=1M_\odot$, $\rho = 10^{-9}$\,g\,cm$^{-3}$, density power law $a=2$, $T_0=600$\,K, temperature power law $m=1$, $\mu=2.25$, $\rho_s = 3$\,g\,cm$^{-3}$ and $\sigma=3.85 \cdot 10^{-15}$\,cm$^2$.}
    \label{fig:weidenschilling}
\end{figure}
The results of these simulations are shown in Figure \ref{fig:weidenschilling}. 
To a high degree of accuracy, the same lifetimes were recovered as were found by \cite{1977MNRAS.180...57W}, differing by less than 3\% on average, with a maximum discrepancy of less than 15\%. This was compared against values extracted from the original plot, using the WebPlotDigitizer tool \cite{WebPlotDigitizer}. This gave us confidence our code was calculating aerodynamic drag correctly across the range of dust sizes and gas densities of interest.

\subsubsection{Analytic Photoevaporative Wind Benchmark}
The final step of our benchmark was to combine both wind and gravitational forces in a photoevaporative wind setting, and simultaneously test our maximum grain size solver \ref{appendix:favourite_section}. \cite{2016MNRAS.463.2725H} was selected for this as they developed an analytic wind model and studied the maximum grain size this wind entrained. The wind model used was derived in \cite{2016MNRAS.461..742H} and is defined as follows:
\begin{align}
    v_{\rm wind} &= c_s \sqrt{-W_0 {\left[-\exp{\left(-\frac{2GM_*}{c^2_s \sqrt{R^2 + z^2}} -1\right)}\right]}}. 
\end{align}
Here $W_0(x)$ is the 0-branch of the Lambert W function and $c_s$ is the gas sound speed.

This assumes an ideal gas and that all quantities only vary in $z$.
We also have to adjust a few previously defined quantities by adding a $z$-component to them. This wind prescription describes an ionisation front, which is defined as the height at which the density drops to below $10^{-5} \times \rho_{\textrm{midplane}}$ \cite{2016MNRAS.461..742H}. This gives us a full description of the disc and wind:
\begin{align}
    \rho(R,z) &= \frac{\Sigma(R,z)}{\sqrt{2\pi} H(R)}\\
    \Sigma(R, z) &= \begin{cases}
        \Sigma_0 (R/R_0)^{-a+1} \exp\left({-\frac{z^2}{2H(R)}}\right) \ \ \mathrm{for} \  \rho \geq 10^{-5} \rho(R,0)\\
        \dot{M} / v_{\rm wind}  \ \ \mathrm{for} \ \rho < 10^{-5} \rho(R,0)
    \end{cases} \label{eqn:sigma_law}\\
    \mu(z) &= \begin{cases}
        2.34 \ \ \mathrm{for} \ \rho \geq 10^{-5} \rho(R,0)\\
        0.88  \ \ \mathrm{for} \ \rho < 10^{-5} \rho(R,0)
    \end{cases}\\
    T(R, z) &= \begin{cases}
        T_0 (R/R_0 )^{-m} \ \ \mathrm{for} \ \rho \geq 10^{-5} \rho(R,0)\\
        10,000K \ \ \mathrm{for} \ \rho < 10^{-5} \rho(R,0)
    \end{cases}\label{eqn:temp_law}
\end{align}
where $\dot{M}$ is the mass loss rate measured at the ionisation front and $H$ is the scale height of the disc. 
We then use the same stellar, T, gas and dust parameters as \cite{2016MNRAS.463.2725H}: $M_*=1M_\odot$, $T_0=629.4$K (equivalent to $H_0=0.05\mathrm{AU}$), $\Sigma_0=100$g/cm$^2$ and $\rho_s = 3$g/cm$^3$. We also select the same range of power laws $a \in [1.0, 2.5]$ and $m \in [0.4,0.8]$. 

The dust was then placed slightly beyond the ionisation front, at the base of the wind and evolved assuming the Epstein regime as \cite{2016MNRAS.463.2725H} did. 

The predicted maximum entrained grain radius is 
\begin{equation}
    s_{\textrm{max, HLM16b}} = \frac{3\sqrt{3}}{2}\frac{c_s\dot{M}R}{\rho_{\textrm{eff}}v_K^2}. \label{eqn:hutch_max_dust}
\end{equation}
Here $v_K$ is the Keplerian velocity and $\rho_{\textrm{eff}}=\rho_s\sqrt{\frac{\pi\gamma}{8}}$ is the same as in \cite{2016MNRAS.463.2725H}, with $\gamma=1.0$.
This equation is derived in detail in \cite{2016MNRAS.463.2725H} by balancing the forces on the dust (i.e. those due to gravity, pressure gradient and the centrifugal force). 

\begin{figure}
    \centering
    \includegraphics[width=\columnwidth]{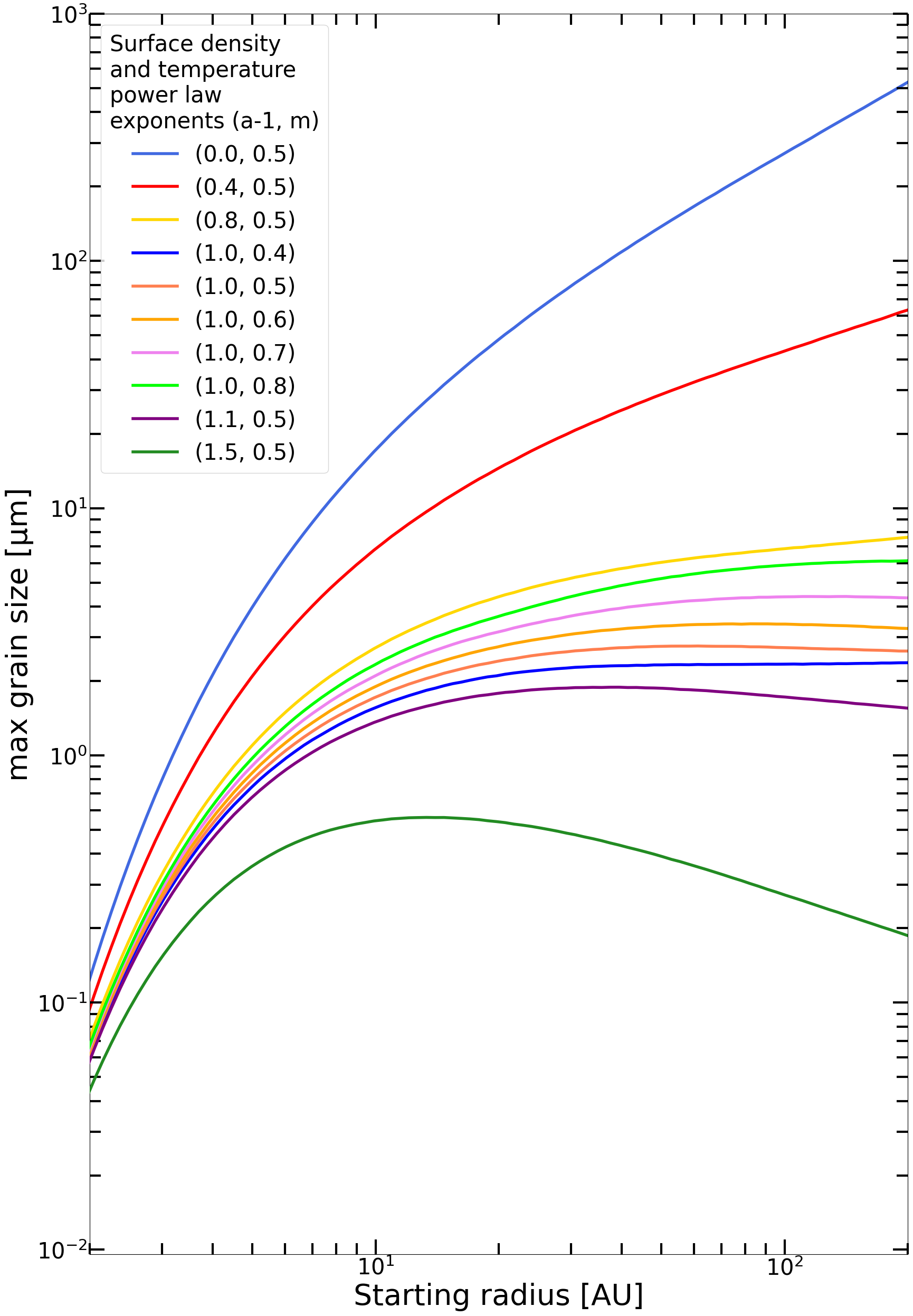}
    \caption{The maximum grain radius entrained in a \protect\cite{2016MNRAS.463.2725H} semi-analytic photoevaporative wind model as a function of the starting radius of the grain. Different lines are for different density $a$ and temperature power laws $m$. (See Eq. \ref{eqn:sigma_law} and Eq.\ref{eqn:temp_law})}
    \label{fig:hutchison}
\end{figure}

We then ran our maximum grain size solver code (Section \ref{appendix:favourite_section}) for the same variety of density and temperature power laws ($a$, $m$) and various starting radii as \cite{2016MNRAS.463.2725H}. We considered the dust ``entrained'' if it reached escape velocity. These results are shown in Figure \ref{fig:hutchison} and are in excellent agreement, on average to within 0.1\% of their results across the entire parameter space where equation \ref{eqn:hutch_max_dust} is valid. 

\section{Results}
\label{sec:results}
We have so far detailed our particle dynamics solver and have tested its performance over long term Keplerian orbital evolution, \cite{1977MNRAS.180...57W} radial drift and the \cite{2016MNRAS.463.2725H} dust entrainment in a planar wind as benchmarks. With the veracity of the code now established we now turn our attention to the new application of dust entrainment in external photoevaporative winds.

\subsection{Dust entrainment in 1D models of external photoevaporation}

External photoevaporation occurs when circumstellar material is irradiated by an external UV source, usually a nearby massive star in a stellar cluster. Ionising extreme ultraviolet (EUV) radiation leads to the characteristic teardrop shaped ionisation fronts of the ``proplyds'' seen in regions such as the Orion Nebula Cluster \citep{2008AJ....136.2136R, 2024arXiv240312604A}. However it is actually typically far ultraviolet (FUV) radiation that launches the wind from the disc \citep{2023MNRAS.526.4315H}. Modelling the launching of such a wind requires the iterative solving of radiative transfer, photodissociation region (PDR) chemistry and hydrodynamics \citep{2016MNRAS.463.3616H}. With the key coolant being the escape of line photons, multidimensional models had historically been prohibitively expensive (though as we will discuss below, 2D axisymmetric models are now possible).  Nevertheless, great progress was made estimating external photoevaporative mass loss rates using 1D calculations. 

\begin{figure}
    \centering
    \includegraphics[width=\columnwidth]{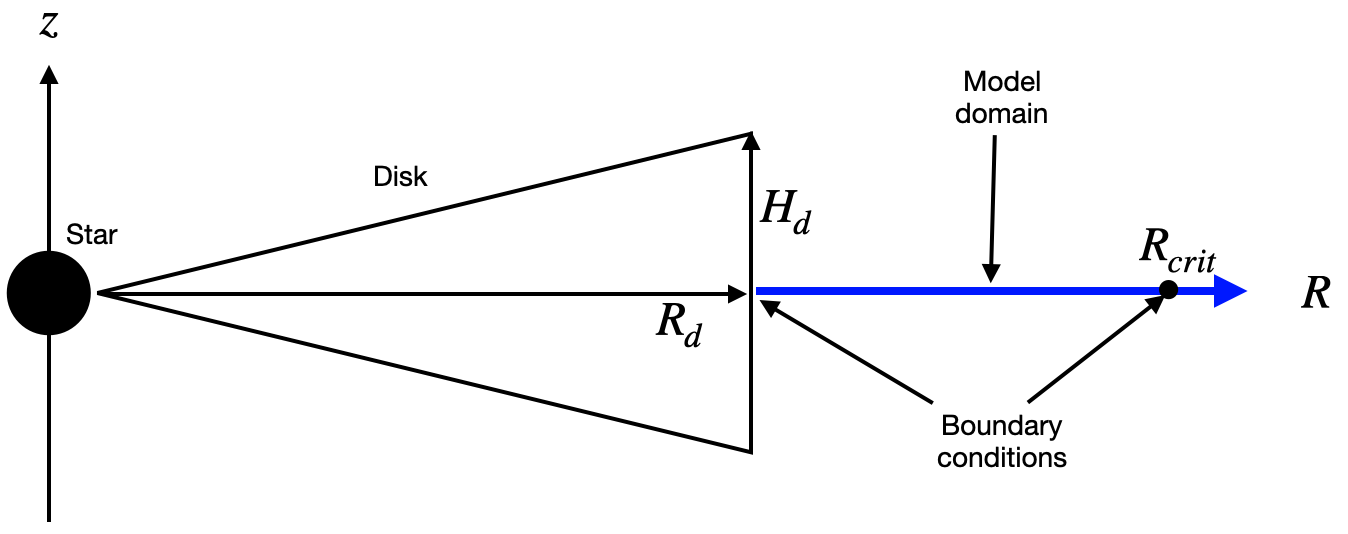}
    \caption{A schematic of the geometry of 1D models of external photoevaporation. The solution domain is from the disc outer midplane. The resulting 1D flow profile is then converted to a ``3D'' mass loss rate by assuming it applies over the entire solid angle subtended by the disc outer edge. }
    \label{fig:1DmodelGeometry}
\end{figure}

\cite{2004ApJ...611..360A} and \cite{2016MNRAS.457.3593F} solved for 1D semi-analytic steady state wind solutions by pre-tabulating the PDR temperature as a function of parameters such as density, column and incident FUV; utilising the disc outer edge and a critical point ($R_\mathrm{crit}$) in the flow as a boundary condition. These models assume that the majority of the mass loss comes from the disc outer edge, where there is a substantial mass reservoir that is most loosely bound to the star. They solve for the density and velocity of the flow from the disc outer edge midplane and assume that flow applies over the entire solid angle subtended by the disc outer edge to calculate a total mass loss rate \citep[][found this to give conservative mass loss rates compared to 2D models]{2019MNRAS.485.3895H}. A schematic of the 1D setup is given in Figure \ref{fig:1DmodelGeometry}. For a disc of radius $R_d$ and scale height at the disc outer edge $H_d$ the solid angle subtended at the disc outer edge is 
\begin{equation}
    \mathcal{F} = \frac{H_d}{\sqrt{H_d^2 + R_d^2}}
\end{equation}
and the 1D profile is converted to a total mass loss rate through
\begin{equation}
    \dot{M} = 4\pi R^2\mathcal{F} \rho v_g,  
    \label{equn:Mdot}
\end{equation}
where $v_g$ is the gas velocity at the base of the wind. This equation is valid at any radial distance $R$.

\cite{2016MNRAS.457.3593F} studied grain entrainment in 1D semi-analytic models of external photoevaporation like those described above, finding that only small grains should be entrained. If grain growth/radial drift has occurred in the disc this then leads to a dust depleted wind, which reduces the shielding of the disc from further UV irradiation and enhances the mass loss rate substantially. They provided an analytic approximation for the maximum entrained grain radius in the limit of a highly subsonic inner wind and no centrifugal force
\begin{equation}
    s_{\textrm{max}} \approx \frac{\bar{v}}{GM_*}\frac{\dot{M}}{4\pi\mathcal{F}\bar{\rho}}
    \label{eqn:amax}
\end{equation}
where $\bar{v}$, $\bar{\rho}$ and $\dot{M}$ are the mean thermal speed of the gas molecules, mean mass density of the dust grain and mass loss rate.  
Given the importance of grain entrainment for the mass loss rates we first apply our particle solver to testing this approximation. We will then move on to studying the entrainment of dust in 2D external photoevaporation calculations where this analytic approximation does not apply at all points of the base of the wind. 

\subsubsection{1D external photoevaporation: Model details}
The 1D calculations described above were steady state semi-analytic models that can only be solved where a critical point in the flow can be calculated as a boundary condition. For our study of we use slightly different PDR-hydrodynamics calculations that can provide a solution depending only on a disc boundary condition. These calculations are computed with the \textsc{torus-3dpdr} code \citep{2015MNRAS.454.2828B, 2019A&C....27...63H} and were used to produce the large publicly available \textsc{fried} grids of mass loss rates \citep{2018MNRAS.481..452H, 2023MNRAS.526.4315H} where the calculations are described in detail. These models cover a wide range of disc sizes ($40\mathrm{AU}$-$500\mathrm{AU}$), central-star masses ($0.1M_\odot$-$3.0M_\odot$), surface densities ($10\Sigma_{0, 1\mathrm{AU}}$-$1000\Sigma_{0, 1\mathrm{AU}}$) and radiation strengths ($100\mathrm{G_0}$-$100,000\mathrm{G_0}$). 

We take the final steady state flow from the 1D FRIEDv2 grid of models \citep{2023MNRAS.526.4315H} and introduce grains of different sizes at the base of the wind without any initial velocity. We assume that the gas properties remain steady, i.e. there is no back-reaction of the dust on the gas, which is reasonable given the expectation is that the dust-to-gas mass ratio in the wind is expected to be depleted. There are then three possible fates of the dust:
\begin{enumerate}
    \item The dust drifts radially inwards back towards the central star. This dust is obviously not entrained.
    \item The dust is dragged out a short way, but reaches force balance before it passes the critical point that \cite{2016MNRAS.457.3593F} use to define their outer boundary. This dust is also classified as not entrained. 
    \item The dust is dragged beyond the critical point of \cite{2016MNRAS.457.3593F} and reaches the escape velocity $v_{\mathrm{esc}}=\sqrt{\frac{2GM_*}{|\vec{r}|}}$ and is classed as entrained. 
\end{enumerate}
The grid is imported into our code, and the values at any given point in the dust evolutionary calculation are linearly interpolated from those. Therefore, our results should be resolution independent as long as the grid is at least as fine as the input model. Some parts of the FRIEDv2 grid considered small discs and/or weak UV fields wherein the flow solution was at best pseudo steady and \cite{2023MNRAS.526.4315H} utilised a time averaging to estimate the mass loss rate in those cases. For the entrainment of dust and comparison with \cite{2016MNRAS.457.3593F} we utilise only those portions of the grid where the velocity profile was purely radially outwards at all radii and not subject to numerical oscillations that were sometimes seen in the pseudo-steady hydrodynamic models. 

We also further filtered out some of the FRIEDv2 models that were deemed unstable based on the following automated criteria: if more than 3 grid points in a row had a negative radial velocity. Otherwise, single negative points were replaced with the average of its neighbouring points, to smooth over numerical oscillations. In the end, 196 different models over the aforementioned parameter range were used.
We then use our dust solver, described in section \ref{appendix:favourite_section}, iteratively solving for the size of entrained dust in each model to a relative precision of 0.1\%.

\subsubsection{1D external photoevaporation: Results}
\begin{figure}
    \centering
    \includegraphics[width=\columnwidth]{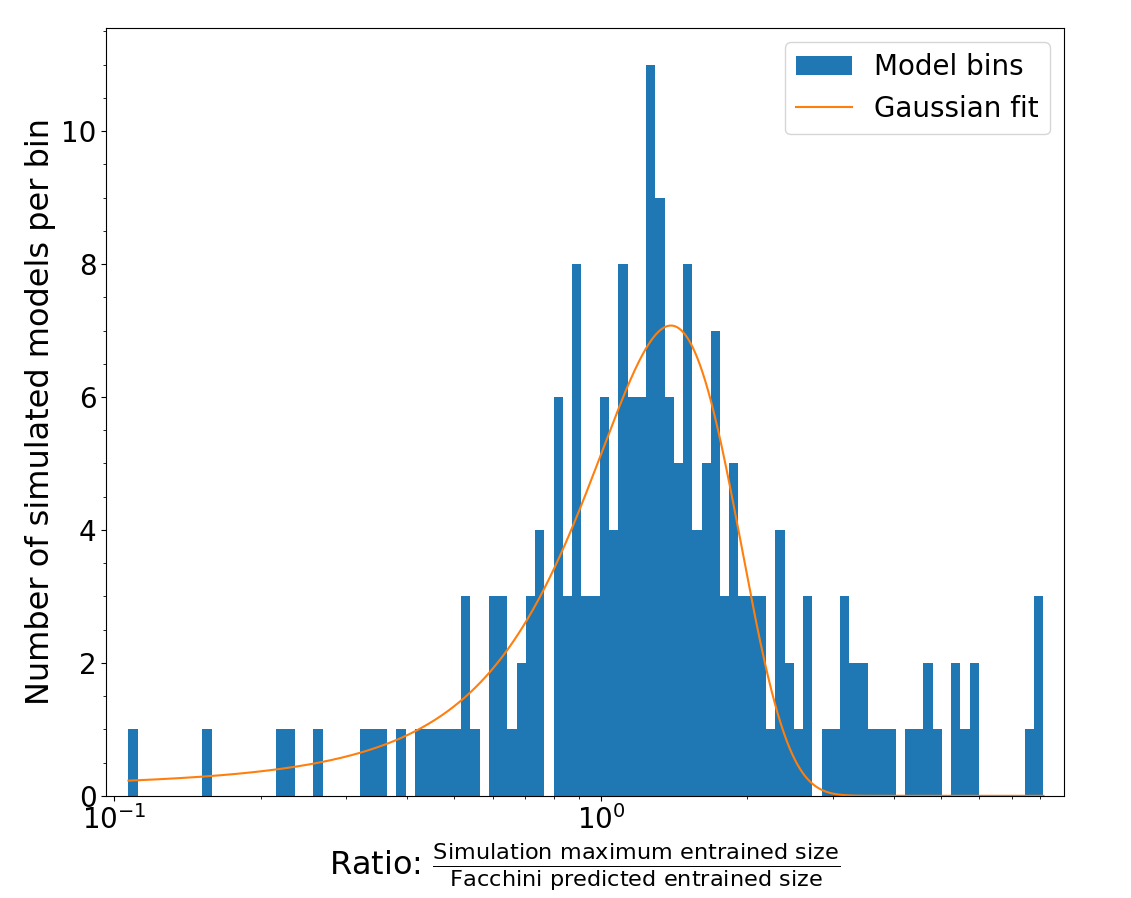}        
    \includegraphics[width=\columnwidth]{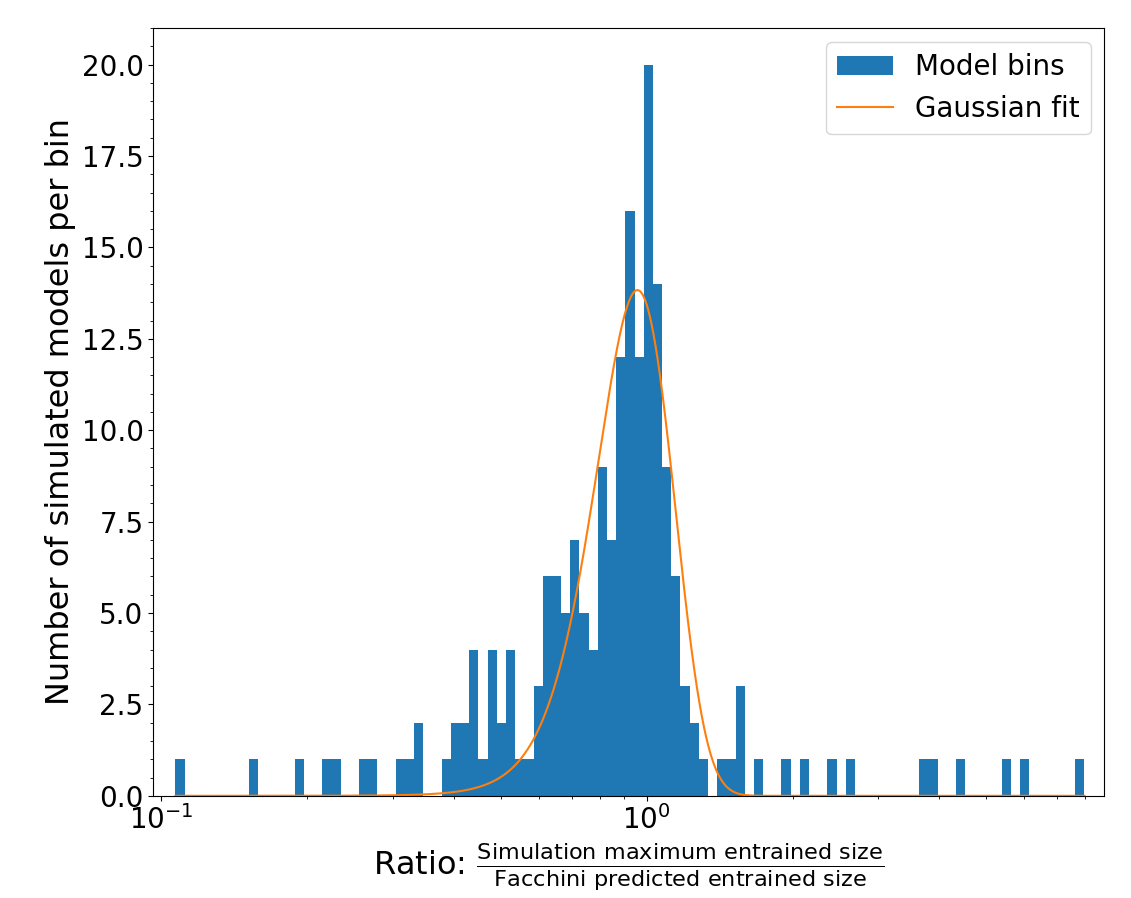}
    \caption{ The upper panel shows the distribution of ratios of the maximum entrained grain size in our models compared to the approximate analytic maximum value from \protect\cite{2016MNRAS.457.3593F}. Our numerical models are typically around a factor 2 higher. The lower panel shows the same distribution for simulations where the centrifugal force is absent, which brings the models and analytic approximation into good agreement. }
    \label{fig:facchini_comparison}
\end{figure}

In Figure \ref{fig:facchini_comparison} are histograms of the distribution of the ratio of the maximum entrained grain size by our code to that expected from the analytic approximation of \cite{2016MNRAS.457.3593F} (equation \ref{eqn:amax}). We reiterate that the analytic approximation assumes that the wind is highly subsonic and ignores the centrifugal term. The upper panel of Figure \ref{fig:facchini_comparison} shows a histogram of the ratio of the two maximum sizes when the dust is started with a Keplerian orbital velocity ($v_\phi=v_K$). There is a spread, but our models typically entrain dust around a factor two larger than the analytic approximation. To understand why the numerical models are systematically entraining larger dust than predicted by the analytic approximation we re-ran the numerical models without the centrifugal term, the results of which are shown in the lower panel  of Figure \ref{fig:facchini_comparison}. There is still a spread in the agreement, but it is now centred on the numerical models being consistent with equation \ref{eqn:amax}. From this we can conclude that the centrifugal force is responsible for the factor of two shift of the maximum entrained dust. The models that show stronger deviation from the expected maximum entrained grain size are due to the underlying radiation hydrodynamic model of the gas flow only being pseudo-steady. The \textsc{fried} calculations only required that the time averaged mass loss rate be steady, but in reality that is sometimes achieved with numerical oscillations in the flow that affect the dust dynamics. 

In summary we find that the analytic approximation for the maximum grain size entrained in external photoevaporative winds by \cite{2016MNRAS.457.3593F} is typically an underestimate of around a factor two, with the difference being explained by the centrifugal force term. The approximation is being used in studies of the dust evolution of externally photoevaporating discs, \citep[e.g.][]{2020MNRAS.492.1279S, 2023arXiv231020214G} so we have validated that approximation for its use in these scenarios, and a factor two correction could be applied to account for the centrifugal contribution in future applications.

\subsection{Dust entrainment in 2D models of external photoevaporation}
We have so far verified that in the 1D framework the maximum grain size entrained in the wind is given by equation \ref{eqn:amax} to within typically a factor 2. However the reality of external photoevaporation is more geometrically complex. There is a decreasing density profile of the disc above the midplane and some components of the wind are also launched from the disc surface. The streamline morphology, density, temperature and velocity can vary substantially for different parts of the wind and all influence the dust dynamics. Existing 2D external photoevaporation models assume that everywhere outside of the disc, the dust is similar to that expected from the 1D models \citep{2019MNRAS.485.3895H}. However it is possible that the dust in the wind might vary substantially as a function of height above the midplane, which in turn could plausibly be expected to affect the attenuation of the FUV field incident upon different components of the disc surface. This would also affect the expected observational characteristics of the dust in the wind. It is therefore important to understand how different grain entrainment is away from the disc outer edge midplane, so we now turn our attention to the unexplored territory of entrainment of dust in 2D external photoevaporation simulations.

\subsubsection{2D external photoevaporation: Model details}
\begin{figure}
    \centering
        \includegraphics[width=\columnwidth]{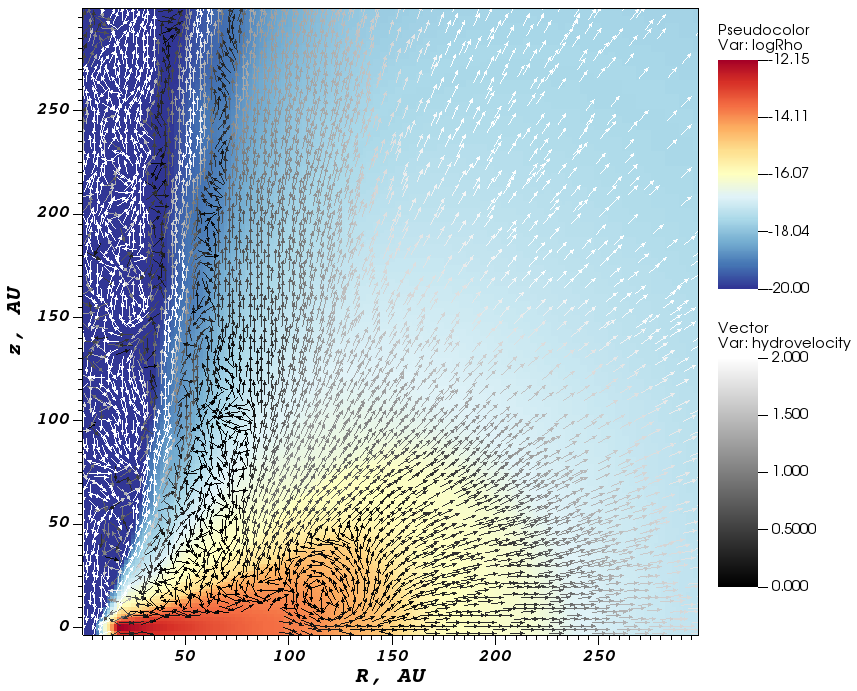}
    \caption{The density and velocity structure near the disc in our 2D radiation hydrodynamic simulation. There is an imposed disc boundary condition (the component of the grid with no velocity vectors) and a wind evolves in the rest of the domain. The flow consists of an inner atmosphere to the disc, associated with low velocity flow of material, including a slow vortex due to shear. Beyond that atmosphere, material streams away more rapidly in a roughly spherically diverging flow.}
    \label{fig:disc_dens_vel}
\end{figure}

We utilise the results of the 2D-axisymmetric steady state external photoevaporation models that appeared in  \cite{2023MNRAS.518.5563B}.  These models are computed on a cylindrical $R-z$ grid. They impose a ``disc'' boundary condition out to some distance $R_d$ and up to one scale height. The steady state wind solution from that boundary was then calculated with a \textsc{torus-3dpdr} radiation hydrodynamic simulation \citep{2015MNRAS.454.2828B, 2019A&C....27...63H}. For our simulations we used a 100\,AU-sized disc in an isotropic $5000\mathrm{G}_0$ radiation environment; the density and velocity distribution of which are given in Figure \ref{fig:disc_dens_vel}. The flow is at large distances from the disc surface is approximately spherically diverging, however closer to the ``disc'' boundary condition things get more complicated. There is a hydrostatic atmosphere where the low velocity flow structure can be complicated, including a (slow) vortex like feature between the disc outer edge and surface. This is due to shear at the corner of the imposed disc boundary and is most likely not physical. 

We again introduce grains of varying size at the base of the wind and study their trajectories over time. Two sets of starting positions were selected: just beyond the imposed disc edge at $R_d$, with height varying from the midplane to one scale height; and at a height of 30\,AU from a radius of 30\,AU to 100\,AU. This height was selected to avoid the stationary wind very close to the disc surface.

\subsubsection{2D external photoevaporation: spatial distribution of dust in the wind}
\label{sec:2Ddustmap}
\begin{figure}
    \centering
    \includegraphics[width=\linewidth]{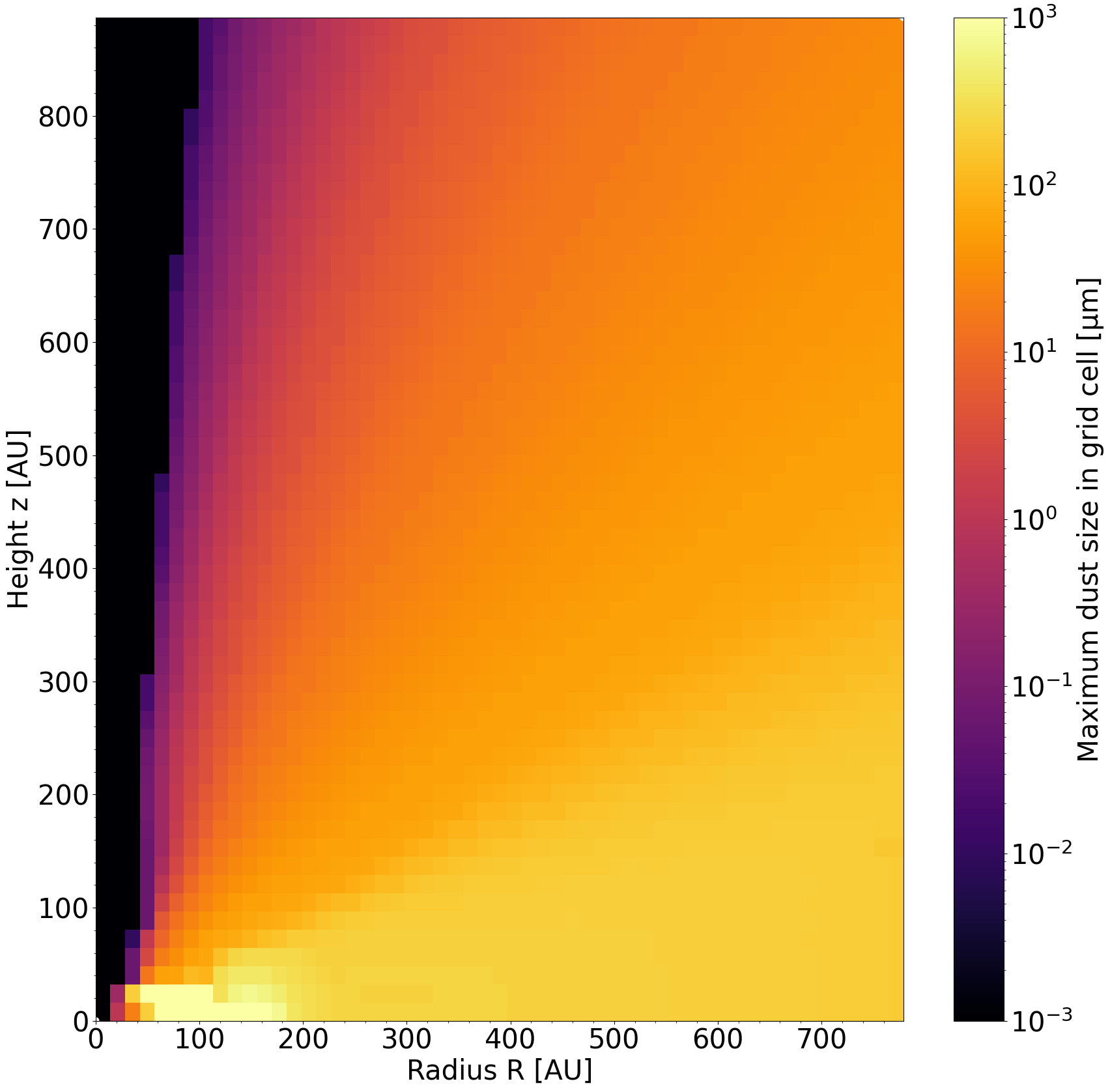}
    \caption{A map of the maximum dust size at any given point in the wind, with a resolution of 15\,AU for the 100\,AU, $5000$G$_0$ 2D-axisymmetric model.}
    \label{fig:2D_dust_map_aentr}
\end{figure}

A map of the maximum grain size throughout the wind in the 100\,AU disc, $5000\mathrm{G}_0$ model is shown in Figure \ref{fig:2D_dust_map_aentr}. The main behaviour is a decrease in maximum entrained grain size as a function of polar angle from the midplane. In the portion of the flow emanating from the disc outer edge, the maximum entrained size is tens to a hundred microns. Whereas at higher polar angle this progressively drops down towards micron (even sub-micron) sized grains. We have demonstrated here that there is indeed a strong variation in the sizes of dust in external photoevaporative winds. A radial gradient of dust sizes in the wind is also observed, where dust hundreds of microns in size stagnates in the wind a few hundred AU away from the disc radius. The maximum size of this stagnant dust decreases as it gets further from the disc edge. This same gradient is not observed for dust launched from the disc surface, where larger dust instead falls back into the disc or into the wind launched from the disc edge.

\begin{figure}
    \centering
    \includegraphics[width=1\linewidth]{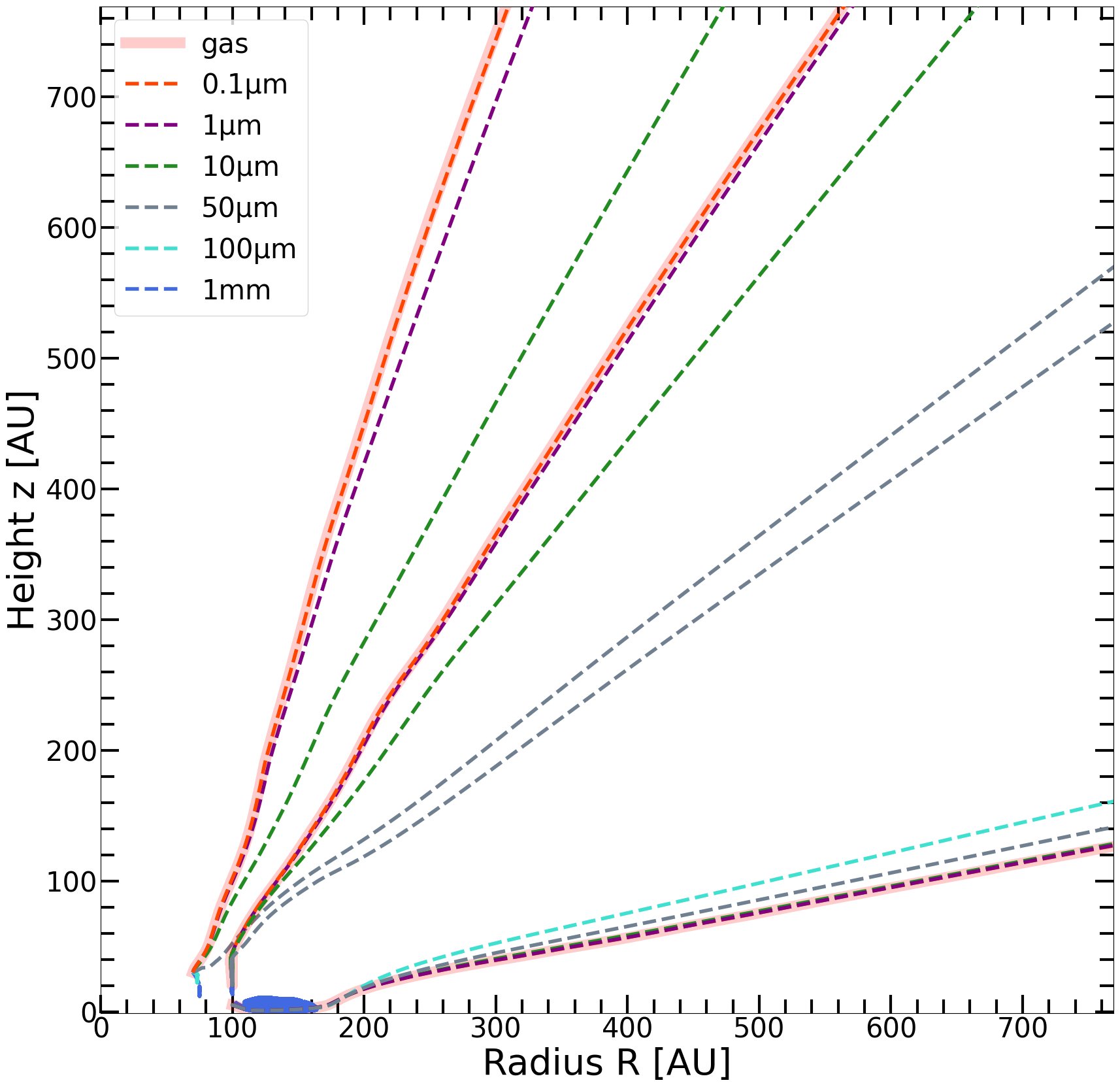}
    \caption{Graph of gas streamlines (translucent line) originating from three different starting points on the disc outer edge or surface. The other (dashed) lines show the paths taken by dust grains of various sizes starting from the same starting points.}
    \label{fig:streamlines}
\end{figure}

We now turn our attention to the trajectories of grains of different sizes, and when they deviate from the gas streamlines. Figure \ref{fig:streamlines} shows the gas streamlines and a collection of dust paths, for different dust sizes and points of origin near the disc surface. Three starting locations were chosen to discuss different behaviours observed: the disc outer edge and two points from the disc surface. There are a range of interesting behaviours. From the disc outer edge of Figure \ref{fig:streamlines}, we see that for a streamline originating from just above the midplane at the disc outer edge, dust of 10\,$\mu$m or smaller closely follows the gas, whereas larger grains are actually launched \textit{above} the gas streamline. This was determined to be due to the larger grains' angular momentum.

The wind is substantially weaker from the disc surface and so the drag force is weaker. In Figure \ref{fig:streamlines} we see that the 50 and 100\,$\mu$m grains therefore simply settle back down into the disc, and for the point of origin from the disc surface even closer to the central star, the 10$\mu$m dust also settles. For the wind from the disc surface, even the small grains (1-10\,$\mu$m) can be substantially dynamically decoupled from the gas, and typically undershoot the gas streamline, being brought back down closer to the midplane by gravity. 

\subsubsection{2D external photoevaporation: Spatial variation of the dust-to-gas ratio and FUV cross section, $\sigma_{\textrm{FUV}}$}

A key quantity set by the dust in the wind is the cross section to the FUV, $\sigma_{\textrm{FUV}}$, which \cite{2016MNRAS.457.3593F} calculate at a reference wavelength of 0.1\,$\mu$m. $\sigma_{\textrm{FUV}}$ is important because it determines the attenuation of the external FUV radiation by dust in the wind \citep[see also][]{2023MNRAS.526.4315H}. \cite{2016MNRAS.457.3593F} studied the behaviour based on the flow from the midplane, but it is previously unexplored how it might vary in a multidimensional flow. We have shown above that the maximum entrainable grain size at any point in the flow can vary substantially, so we now evaluate the impact that would have on the value of $\sigma_{\textrm{FUV}}$. Our models did not include dust vertical settling or radial drift calculations, so here we only consider an upper bound on the dust-to-gas ratio in the wind and the resulting FUV opacity it creates.

\begin{figure}
    \centering
    \includegraphics[width=\linewidth]{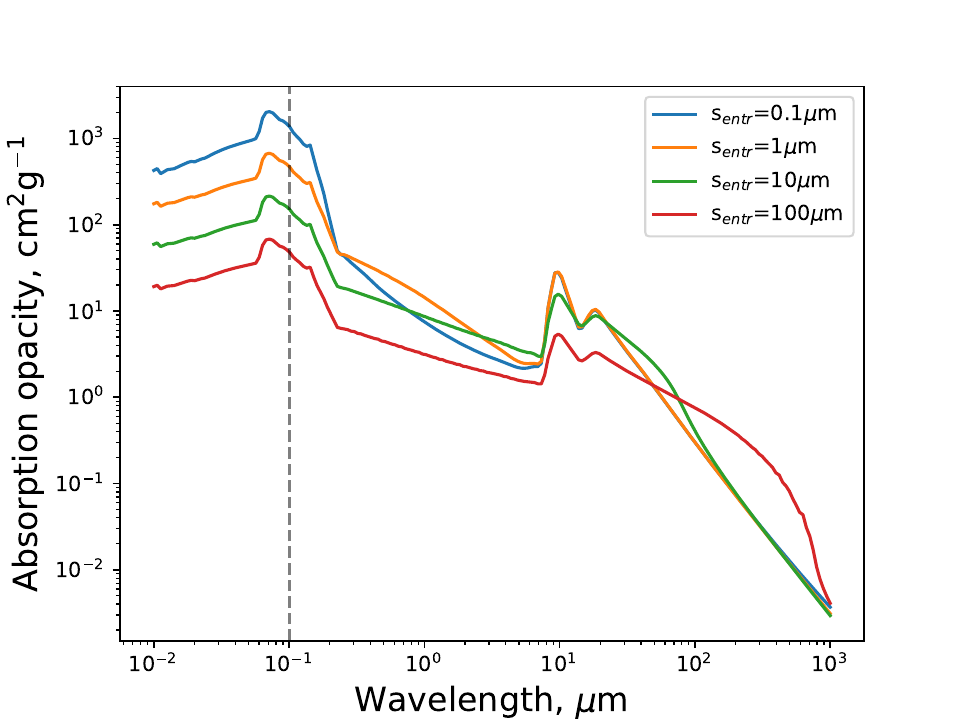}
    \caption{The absorption opacity as a function of wavelength for Draine silicates with size distribution power law scaling as $q=-3.5$. Different lines are for different maximum grain sizes of the grain size distribution (which we refer to as $s_\textrm{entr}$ since locally that maximum size will be the largest entrained grain size). The vertical dashed line is 0.1$\mu$m, at which we calculate $\sigma_{\textrm{FUV}}$.  }
    \label{fig:opacities}
\end{figure}

\begin{figure}
    \centering
    \includegraphics[width=\linewidth]{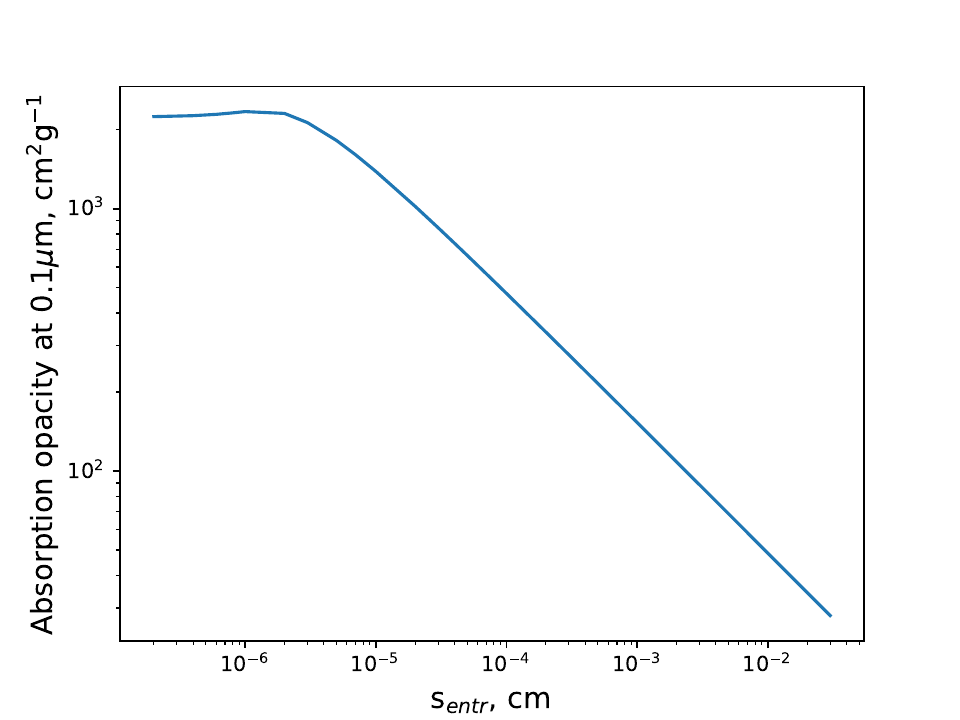}
    \caption{The absorption opacity per gram of gas at $\lambda=$0.1$\mu$m as a function of local maximum entrained dust size. This is used in the calculation of the cross section to the UV, $\sigma_\textrm{FUV}$, see equation \ref{eqn:sigmaFUV}. }
    \label{fig:opacitywithsentr}
\end{figure}

To calculate $\sigma_{\textrm{FUV}}$, we make the assumption that at any point in the flow the grain size distribution holds the same power law as in the disc, up to whatever the maximum entrained grain size is at that point in the flow. The FUV cross section is
\begin{equation}
    \sigma_{\textrm{FUV}} = \mu m_H \kappa(s_{\textrm{entr}}) \frac{\delta_{\textrm{d2g}}(s_{\textrm{entr}})}{100} 
    \label{eqn:sigmaFUV}
\end{equation}
where $m_H$ is the molecular mass of Hydrogen, $\kappa(s_\textrm{entr})$ is the opacity at 0.1\,$\mu$m for a grain size distribution up to the local max entrained grain size and $\delta_\mathrm{d2g}(s_\mathrm{entr})$ is the dust-to-gas mass ratio at the local point in the flow. Following \cite{2016MNRAS.457.3593F}, we define this ratio as 
\begin{equation}
    \delta_{\textrm{d2g}} = 100 \left(\frac{s_{\textrm{entr}}^{4-q} - s_{\textrm{min}}^{4-q}}{s_{\textrm{max}}^{4-q} - s_\textrm{min}^{4-q}}\right)
    \label{eqn:d2g_rat}
\end{equation}
where $s_{\textrm{max}}$ is the global maximum grain size (largest grain size in the disc and at the base of the wind), $s_{\textrm{entr}}$ is the maximum grain size at a given point in the flow, $s_{\textrm{min}}$ is the global minimum grain size and $q$ is the power law of the grain size distribution. For all the following, we set the minimum dust size to $s_\mathrm{min}=2\mathrm{nm}$. In the ISM the fiducial value of $q$ is 3.5, however grain growth in discs typically results in $3.0<q<3.5$ \citep[e.g.][]{2014prpl.conf..339T}. 
We use the \textsc{torus} code \citep{2019A&C....27...63H} to calculate the absorption opacity using Mie theory and optical constants from \cite{1984ApJ...285...89D}. We do so for a range of maximum grain sizes and assuming a fiducial power law slope of the size distribution $q=3.5$ (though we also consider $q=3.0$ for reference). Some examples of the resulting absorption opacity as a function of wavelength in the $q=3.5$ case are shown in Figure \ref{fig:opacities}. Note that \textsc{torus} calculates the opacity per gram of gas and in computing these distributions we assume a gas-to-dust mass ratio of 100 (modifications due to the dust-to-gas ratio are introduced in equation \ref{eqn:sigmaFUV}). We then tabulate the opacity at 0.1$\mu$m as a function of maximum grain size from the set of opacity models which we use to interpolate opacities for arbitrary maximum entrained grain sizes in the dust entrainment calculations, as illustrated in Figure 
\ref{fig:opacitywithsentr}. 

\begin{figure}
    \centering
    \includegraphics[width=0.87\linewidth]{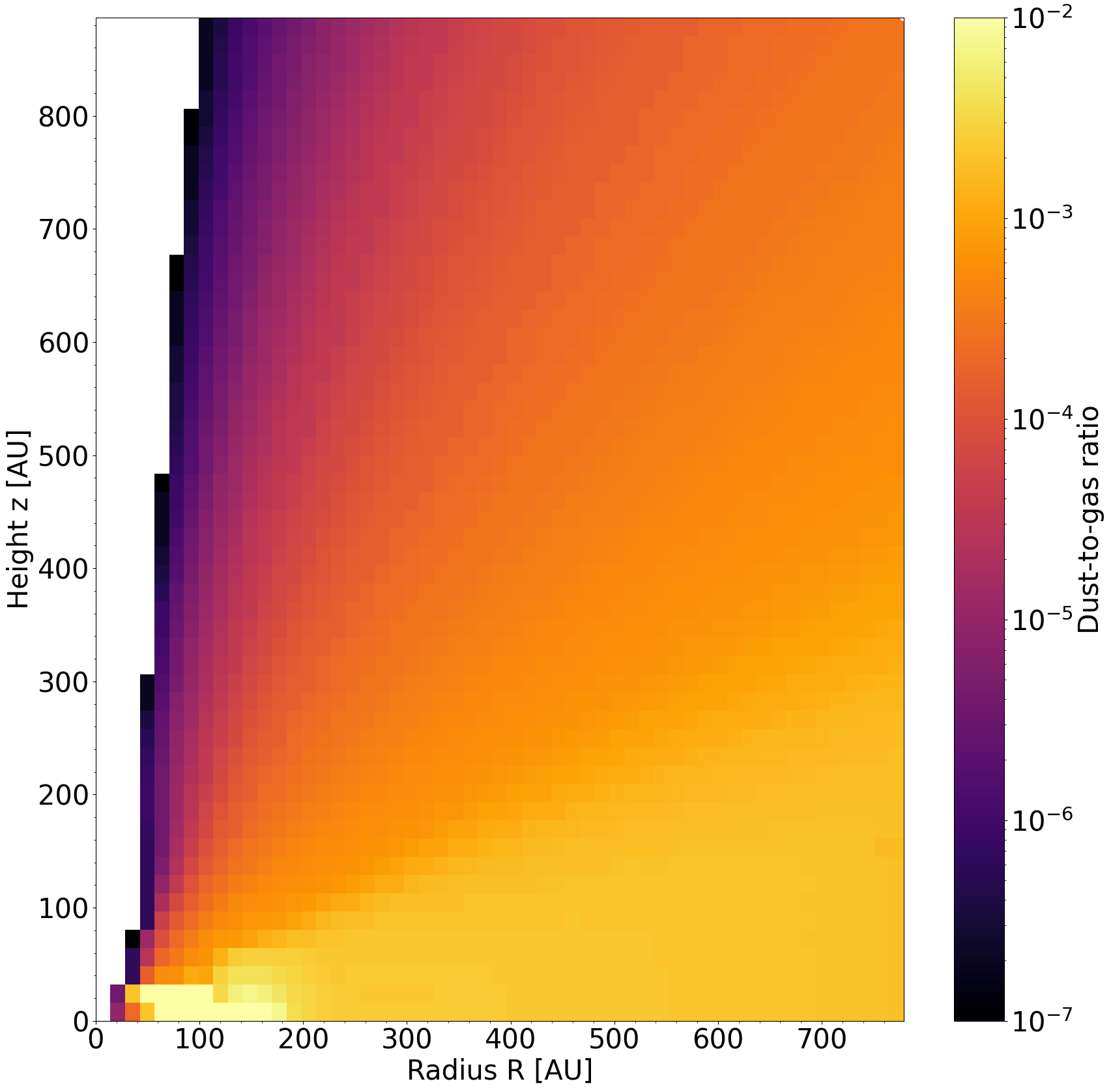}
    \includegraphics[width=0.87\linewidth]{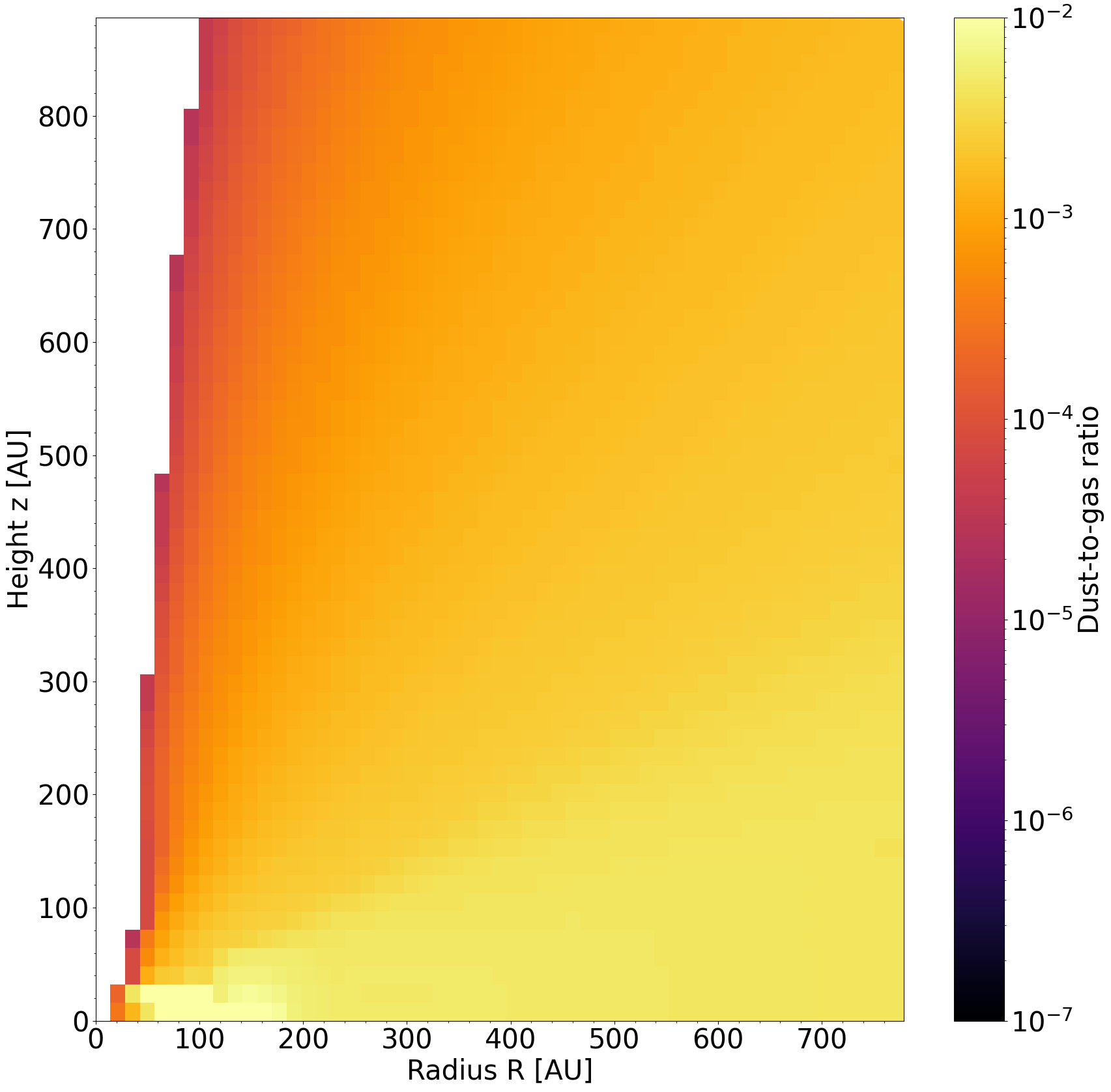}
    \includegraphics[width=0.87\linewidth]{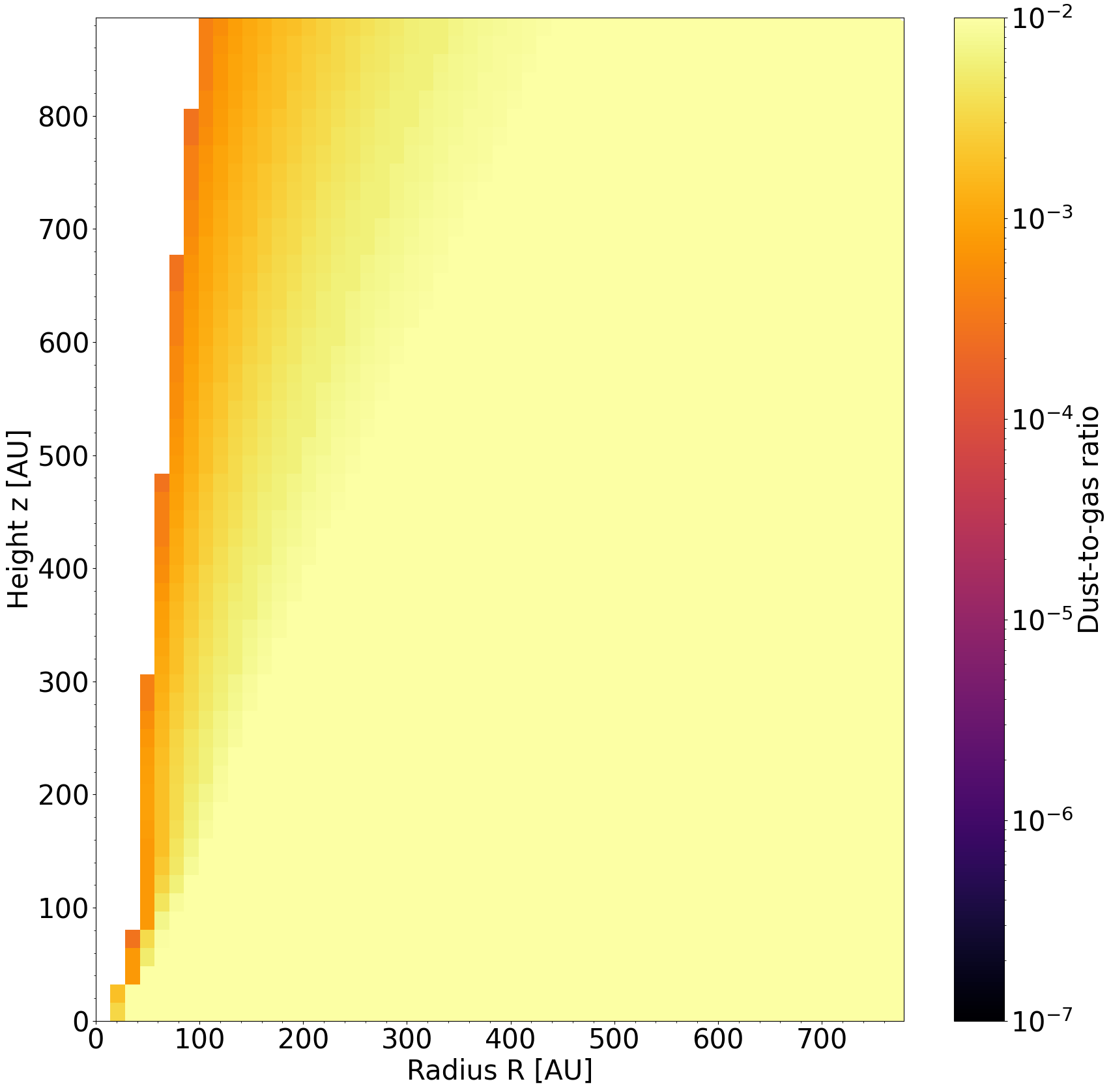}    
    \caption{2D maps of the dust-to-gas mass ratio in the wind for different grain size power law distributions and maximum grain sizes. From top to bottom the panels have the following parameters: $q=3.0$ and $s_\mathrm{max}=1\mathrm{mm}$, $q=3.5$ and $s_\mathrm{max}=1\mathrm{mm}$; and $q=3.5$ and $s_\mathrm{max}=1\mu\mathrm{m}$.}
    \label{fig:d2gmap}
\end{figure}

As discussed in section \ref{sec:2Ddustmap}, we find that the maximum dust size in the wind has a large angular variation. From $1 \mathrm{mm}$ near the midplane at the disc edge, down to sub-micron from the disc surface near the star (as shown in Figure \ref{fig:2D_dust_map_aentr}).  This directly impacts the dust-to-gas mass ratio maps, three examples of which are given in Figure \ref{fig:d2gmap}. The three panels consider grain size power law distributions of $q=3.0$ and $q=3.5$, and also maximum grain sizes in the disc ($s_\mathrm{max}$) of $1\,$mm and $1\,$$\mu$m. In the scenario with $s_\mathrm{max}=1\,\mu$m the dust-to-gas ratio in the wind is fairly uniformly ISM-like at $\delta_{\textrm{d2g}}=10^{-2}$. In the $q=3.5$ and $s_\mathrm{max}=1$\,mm case there is roughly order of magnitude depletion of the dust-to-gas ratio in the wind from the outer edge and stronger depletion as a function of polar angle. In other words, as the maximum grain size in the disc decreases, so does the angular variation of dust in the wind.
The other parameter we can vary is the dust power law - looking at $q=3.0$, we can consider the other extreme. Considering the dust-to-gas ratio at $q=3.0$, with an $s_\mathrm{max}=1\mathrm{mm}$, we can see that with a smaller $q$, the dust-to-gas ratio varies significantly more with angle than in the $q=3.5$ (more than three orders of magnitude!). In other words, as the power law decreases, the angular variation of the dust in the wind increases sharply. 
These effects can be seen by looking at equation \ref{eqn:d2g_rat}: a smaller $q$ leads to a larger power for each term, amplifying the change in $\delta_{\textrm{d2g}}$ as $s_{\textrm{entr}}$ changes in the wind. On the other hand, decreasing $s_\mathrm{max}$ reduces the overall variation in the wind, with more of the wind capping out at $\delta_\mathrm{d2g}=0.01$.

\begin{figure}
    \centering
    \includegraphics[width=0.82\linewidth]{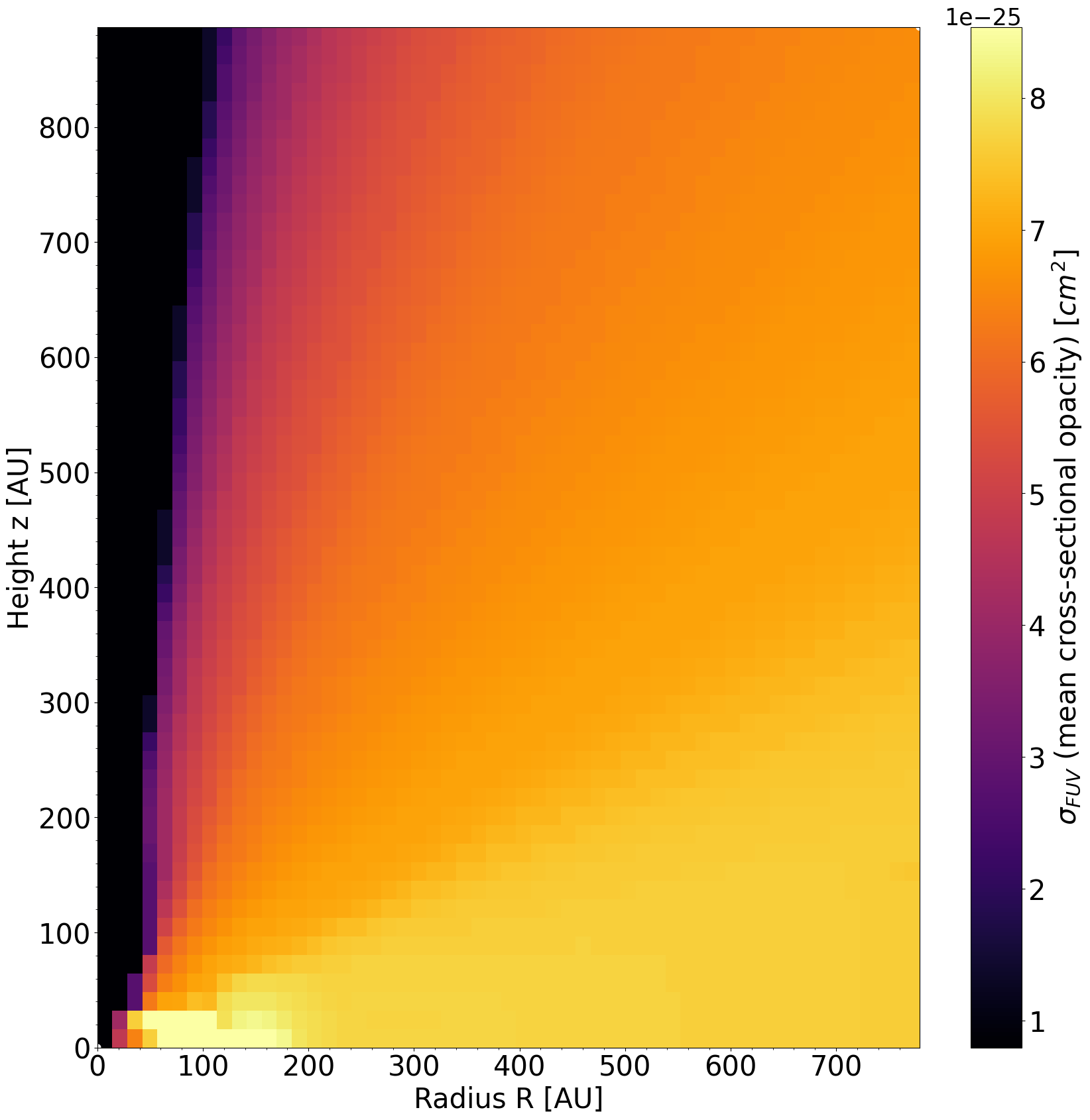}
    \includegraphics[width=0.82\linewidth]{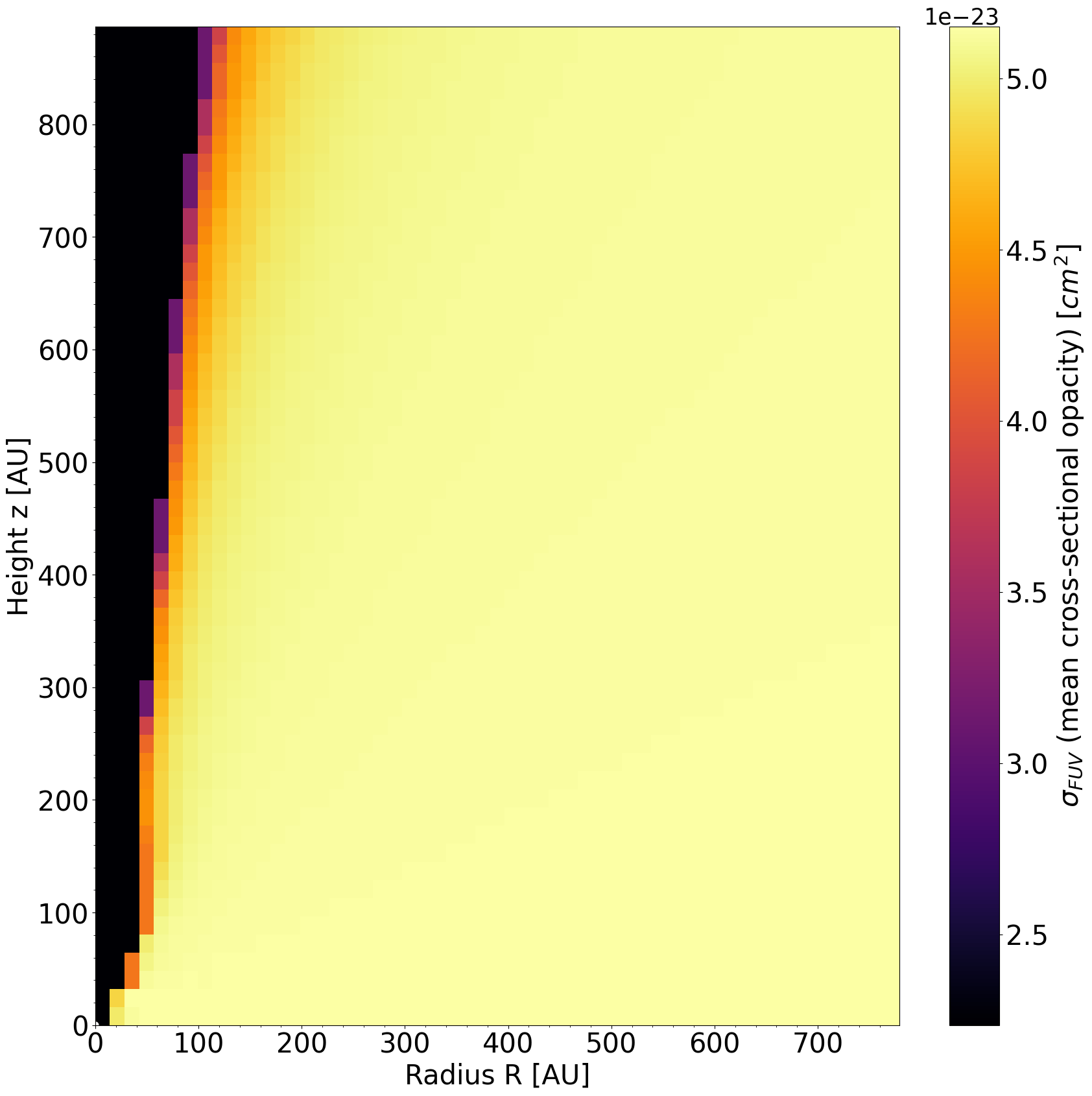}
    \includegraphics[width=0.82\linewidth]{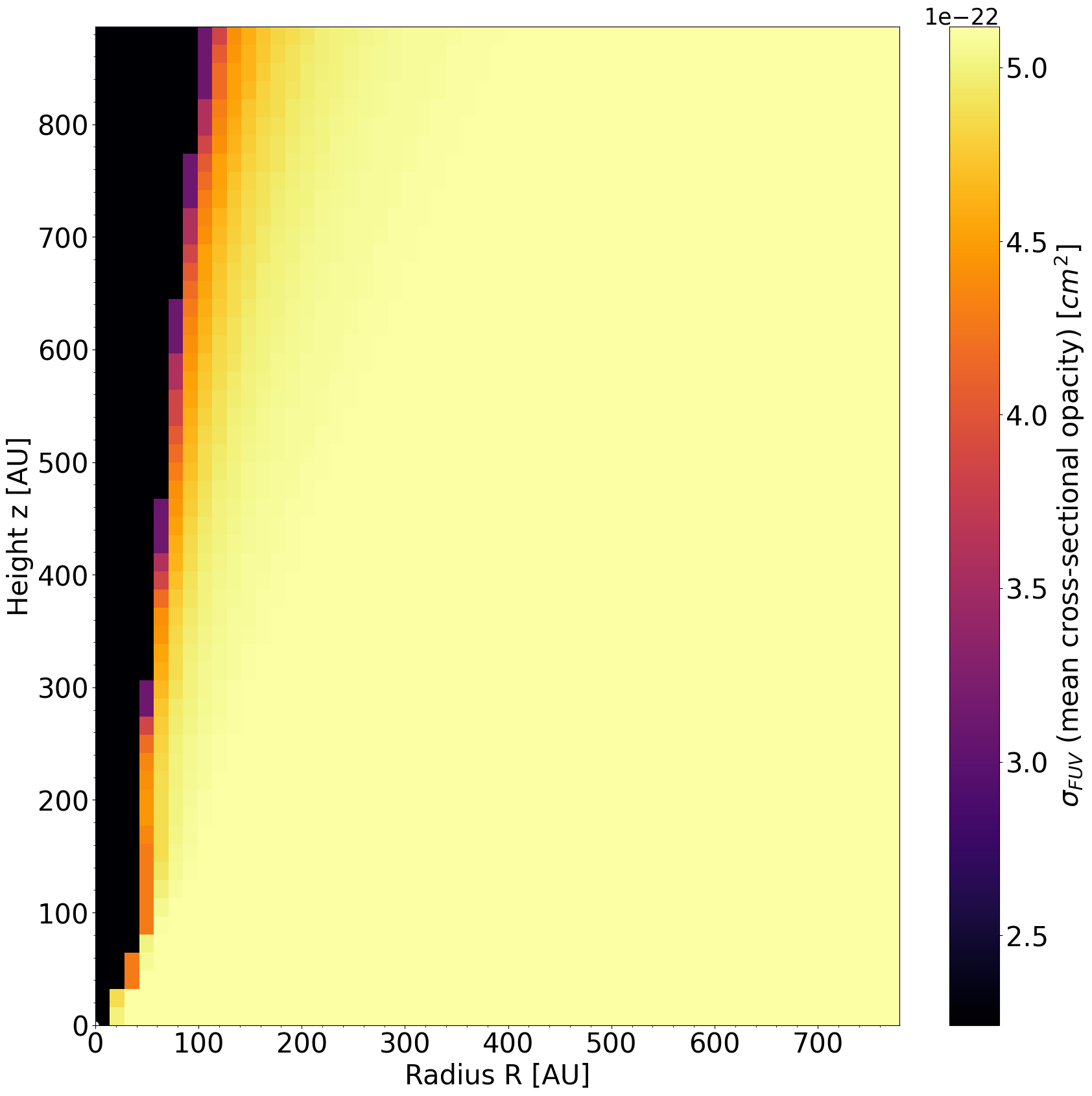}
    \caption{2D maps of the $\sigma_{\textrm{FUV}}$ opacity in the wind for different grain size power law distributions and maximum grain sizes. From top to bottom the panels have the following parameters: $q=3.0$ and $s_\mathrm{max}=1\mathrm{mm}$, $q=3.5$ and $s_\mathrm{max}=1\mathrm{mm}$; and $q=3.5$ and $s_\mathrm{max}=1\mu\mathrm{m}$. (Note the colour axes have different scales.)}
    \label{fig:sigmaFUV_map}
\end{figure}

Three 2D maps of cross sections to the UV, $\sigma_\mathrm{FUV}$, for the same parameters are given in Figure \ref{fig:sigmaFUV_map}. While the models show a large angular variation of entrained dust sizes - three orders of magnitude between what is entrained at the midplane and what is entrained nearer the star at the disc surface - the resulting cross section to the UV is mostly uniform from all directions. This is because the smallest dust is responsible for the majority of the shielding from radiation. Expanding $\sigma_\mathrm{FUV}$ as a function of $s_{\mathrm{entr}}$ we find that the cross section is mainly dependent on the entrained and minimum dust sizes:
\begin{align}
\kappa(s_\mathrm{entr}) &= \kappa_0 s_\mathrm{entr}^{-n}\\
    \sigma_\mathrm{FUV} &= \mu m_H \kappa_0 s_\mathrm{entr}^{-n} \frac{s_\mathrm{entr}^{1/2} - s_\textrm{min}^{1/2}}{s_\textrm{max}^{1/2} - s_\textrm{min}^{1/2}}.
\end{align}
The opacity power law $n$ is recovered empirically from the \textsc{torus} code (e.g. see Figure \ref{fig:opacitywithsentr}). 
Assuming that $s_\textrm{max} \gg s_\textrm{min}$, we can simplify this into our final expression:
\begin{align}
    \sigma_\mathrm{FUV} &\simeq \frac{\kappa_0\mu m_H}{s_\textrm{max}^{1/2}}\left(s_\mathrm{entr}^{1/2-n} - s_\mathrm{entr}^{-n}s_\textrm{min}^{1/2}\right).
\end{align}
For the $q=3.5$ case, we get $n=0.5$ and the first $s_\mathrm{entr}$ cancels out. From this, we can see that only dust within an order of magnitude or two of the minimum dust size strongly contributes to the shielding. Since even the $1\mu$m dust is larger than the minimum dust size of 2\,nm, we end up with a uniform $\sigma_\mathrm{FUV}$. 
In the case of $q=3.0$, we get an $n=0.85$, giving us a variation of roughly factor two over the entire grid.
So while the absolute magnitude of the shielding calculations are quite sensitive to the chosen dust distribution power law $q$, as well as the minimum dust size, to a lesser extent, the amount of angular variation is not. The maximum dust size does not affect the angular variation of $\sigma_\mathrm{FUV}$ and affects the dust power law $q$ only slightly. Finally, while we have not considered the effects of dust radial drift or vertical mixing, which could substantially lower the dust-to-gas ratio at the base of the wind, these effects would slightly lower the overall opacity but not change the uniformity of the opacity, since this is mostly dominated by the small grains which would still be in the wind. This is an important result, since for future radiation hydrodynamic modelling of externally irradiated discs a single representative value of $\sigma_\textrm{FUV}$ can be used in the flow, rather than having to include live decoupled dust-gas dynamics to accurately solve for the extinction to the disc and hence flow structure.

\section{Discussion}

\subsection{Expected impact of an ionisation front}
We have studied the entrainment of dust in external photoevaporative winds from discs irradiated purely by photodissociating FUV radiation. For many externally photoevaporating discs there will also be an ionizing (EUV) component to the incident radiation that can result in an ionization front. The ionisation front will be associated with a step change in temperature (e.g. from $\sim10^3$\,K to $\sim10^4$\,K) and an associated jump in velocity and reduction in density. While we do not model this fully here we comment on the expected change in drag across the ionisation front. The ratio of Epstein drag force terms just interior to and exterior to the ionisation front (the PDR, subscript $\rm i$ and ionised gas, subscript $\rm ii$) assuming no instantaneous acceleration of the dust across the front is
\begin{equation}
    \frac{F_\textrm{i}}{F_{\textrm{ii}}} = \frac{\rho_{\rm i} c_{\rm i} (v_{\rm i}-v_{\rm dust})}{\rho_{\rm ii} c_{\rm ii} (v_{\rm ii}-v_{\rm dust})}. 
  \end{equation}
where $\rho_i$, $c_i$ and $v_i$ are the density, sound speed and gas velocity in the PDR and $v_{\rm dust}$ is the dust velocity. The mass flux $\rho v$ is conserved, hence
\begin{equation}
    \frac{F_{\rm i}}{F_{\rm ii}} = \frac{c_{\rm i} (1-v_{\rm dust}/v_{\rm i})}{c_{\rm ii} (1-v_{\rm dust}/v_{\rm ii})}. 
\end{equation}
There is a factor $\sim$few change in sound speed across the ionisation front (from $\sim3$ to $\sim10$\,km\,s$^{-1}$). If the gas were also travelling at the sound speed and the grain velocity were say 1\,km\,s$^{-1}$ then there would be a factor 4.5 increase in the drag force across the ionisation front. While we have not explicitly modelled the ionisation fronts in this paper, we do therefore expect that dust reaching any such ionisation front will then experience an enhanced drag relative to that it felt in the PDR, and so the inclusion of ionisation in our models would only enhance entrainment.  






\subsection{Initial comparison with observations}
To date there are only limited studies of the dust entrained in external photoevaporative winds. \cite{2012ApJ...757...78M} used Hubble Space Telescope (HST) observations to study the very large, silhouette disc 114-426 towards the Orion Nebula Cluster. They concluded that the likely external radiation field incident upon the disc is of order $10^2$\,G$_0$. 114-426 exhibits possible warping and evidence for an outer spiral (though follow up observations, for example with ALMA, are still lacking to study that in more detail). By comparing the flux in nebular lines with the ambient H\,\textsc{ii} region \cite{2012ApJ...757...78M} were able to constrain a representative grain size in translucent parts of the disc. They determined that that representative grain size decreased with distance from the central star. A similar trend was recently inferred for the same system using JWST data by \cite{2024arXiv241204356B}. \cite{2021MNRAS.508.2493O} included a dust model in their isothermal 1D slim disc models of external photoevaporation, finding that such a gradient in grain size does arise.

\begin{figure}
    \centering
    \includegraphics[width=1.\linewidth, angle=0]{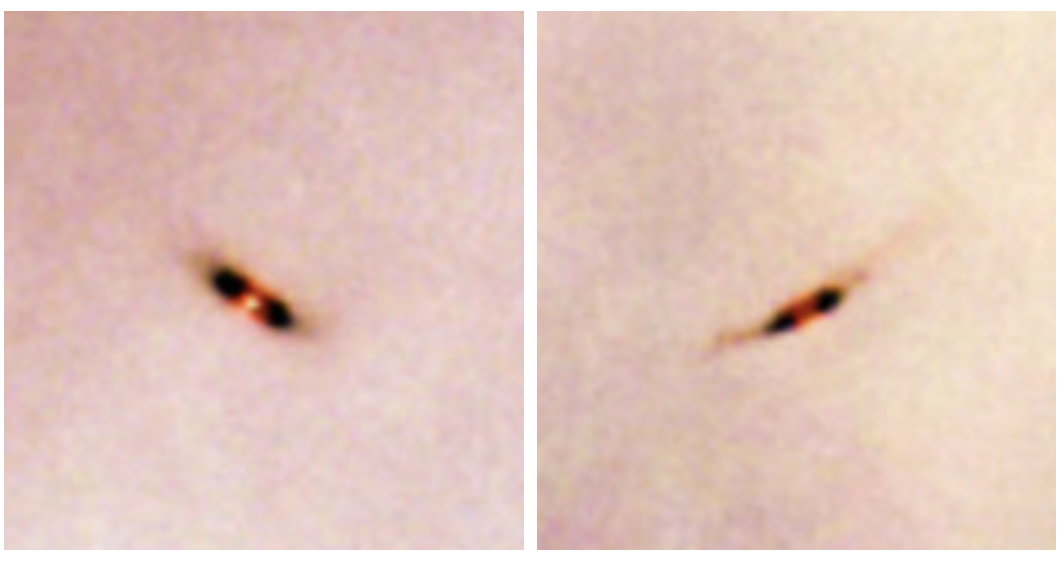}
    \caption{Examples of silhouette discs in the ONC showing evidence of absorption by dust emanating from the disc outer midplane. Reproduced from the press release based on the paper by \protect\cite{2008AJ....136.2136R}. Image credit: NASA, ESA and L. Ricci (ESO).}
    \label{fig:silhouettediscs}
\end{figure}

Our models show that same gradient discussed above, however in 2D we see that it only arises for material emanating from the disc outer edge and stalls (see Figure \ref{fig:2D_dust_map_aentr} and related discussion). Above the disc there is no such gradient with spherical distance, only with polar angle. One would therefore expect the extinction to be greater near the disc outer edge midplane. While a detailed comparison of models and observations using synthetic observations is beyond the scope of this paper, we note that silhouette discs in the ONC do show evidence of this behaviour, with some examples from \cite{2008AJ....136.2136R} given in Figure \ref{fig:silhouettediscs}. \cite{2024Sci...383..988B} also find an extended dusty lobe from the disc outer edge of the 203-506 system. So model predicting stalled dust only from the disc outer edge appears qualitatively consistent with observations. 

We note that the entrained larger dust will likely be difficult to observe in emission (e.g. with ALMA). The density in the wind is at least a few orders of magnitude less dense than the disc (Figure \ref{fig:disc_dens_vel}) and the dust-to-gas ratio is also typically depleted by a further order of magnitude (Figure \ref{fig:d2gmap}). So even if the dust were a factor 10 warmer than in the disc the flux due to dust is likely to be a few orders of magnitude weaker in the wind than in the disc. We will explore simulated observations of our models and the best ways to detect and analyse the entrained dust distribution in future work.

\subsection{Limitation of introducing dust at a fixed disc boundary condition}
We have studied which grain sizes are entrained in steady state external photoevaporative wind models by placing different sized grains at the base of the wind. Here we discuss some limitations of this approach. 

The first caveat is that we do not track the delivery of dust from the disc to the wind. Larger grains settle towards the midplane, and both  \cite{2021MNRAS.502.1569B} and \cite{2021MNRAS.501.1127H} found that for internal photoevaporative winds (launched from the disc surface) the actual limitation on the maximum entrained grain size in the wind was the maximum size that could be delivered from the disc to the wind base. For example \cite{2021MNRAS.501.1127H} find that the maximum entrainable grain size for dust placed at the base of the wind is an order of magnitude higher than the maximum grain size able to reach the wind, though this difference may be partially reduced by turbulent diffusion or some other dredging process. Note that \cite{2024A&A...687A.275J} found the vertical mixing timescale is short, so the wind base should be replenished quickly. For external photoevaporation, entrainment from the disc outer midplane should not suffer this issue unless grain growth and radial drift has depleted the disc dust reservoir in the midplane to only include sizes larger than the maximum entrainable size. However in the outer disc above the midplane and particularly on the outer surface layers the issue of dredging dust up to the base of the wind will be even more significant than for internal photoevaporative winds. The dust above the midplane discussed in the 2D models above may therefore be even more depleted/small than inferred from our models. 

The other caveat is that the steady state wind models that we evolved our dust within define the disc as a boundary condition. In reality there will be a smooth transition from disc to wind and this was studied in isothermal models using a slim disc approach by \cite{2021MNRAS.508.2493O}. This transition will be important again for studying the delivery of dust to the wind base, but since we are not considering that here we retain the use of the fixed boundary conditions. In future, to study grain delivery we would use time evolving simulations that do not solve for a steady state and hence do not utilise a ``disc'' boundary condition. 

Finally, by introducing dust at the disc boundary, we have imposed a fixed and uniform $\delta_{d2g}=0.01$ at the base of the wind for simplicity. In reality, vertical settling occurs on much faster timescales than radial drift \citep[e.g.][]{2004Dullemond&Dominik, 2018Krijt+}, which could lead to a larger dust-to-gas ratio at the disc edge than at the disc surface. This could create an angular gradient of dust-to-gas ratio in the wind, from the midplane up to above the disc. The resulting FUV opacity, however, is still expected to be relatively uniform across the whole wind, as this is dominated by the smallest dust, which would still be present everywhere at the base of the wind.

\subsection{The possible impact of radiation pressure}
One effect not considered in our models is radiation pressure acting upon the dust. An upper limit on the radiation pressure force per unit area assumes that there is zero interaction of the light with material until it encounters completely optically thick disc wind material \citep{2009MNRAS.397.1314N, 2015MNRAS.448.3156H} i.e. 
\begin{equation}
    P_{\textrm{rad}} = \frac{L_*}{4 \pi c D^2}
\end{equation}
where $L_*$ is the luminosity of the radiation source at a distance $D$ and $c$ is the speed of light. The radiation pressure force on a grain is hence 
\begin{equation}
    F_{\textrm{rad}} = \frac{L_*}{4 \pi c D^2}\pi s_{\rm grain}^2. 
\end{equation}
Comparing this to the typical drag force in the wind the radiation pressure is about a factor 100 weaker than wind drag for disc/wind at 0.1\,pc from a $\theta^1$ Ori C like object. At 0.01\,pc from $\theta^1$ Ori C the radiation pressure force would become comparable to the wind drag, but there is only a very small number of proplyds in Orion at that sort of projected separation \citep[e.g.][]{2008AJ....136.2136R}.  We hence expect that radiation pressure is generally of secondary importance for the dust dynamics of externally irradiated discs in regions like Orion, but may play a role in very close proximity to very massive stars.  

\section{Summary and conclusions}

We present a detailed study of the dust entrainment and dynamics in external photoevaporative winds. We developed and benchmarked a new dust dynamics solver and used this to evolve dust within the steady state solutions of 1D and 2D radiation hydrodynamic simulations of external photoevaporation computed with the \textsc{torus-3dpdr} code. We draw the following main conclusions from this work: \\

\noindent 1. We validated the analytic approximation for the maximum entrained grain size from the disc outer edge by \cite{2016MNRAS.457.3593F}. Demonstrating that neglected terms like the centrifugal force introduce around a factor 2 underestimate of the true maximum entrained grain size, but the approximation is accurate enough to be utilised in applications. \\

\noindent 2. We introduce the first models of dust in external photoevaporative winds in 2D. The maximum grain size entrained varies with polar angle. We find $\sim100\,$$\mu$m grains entrained from the disc outer edge, and that maximum size decreases for the weaker flow above the disc down to micron, and even sub-micron sizes. This has a corresponding impact on the spatial distribution of the dust-to-gas mass ratio, with a larger dust abundance in the parts of the flow emanating from the disc outer edge. This bears resemblance to silhouette discs in Orion that appear to have translucent trails of dust from their outer edge in the optical.   \\

\noindent 3. Despite the spatial variation in the dust properties, the cross sectional opacity to the FUV $\sigma_{\textrm{FUV}}$ (defined at 0.1\,$\mu$m) is relatively uniform in the wind. This is because the FUV opacity is predominantly set by the small dust, which is effectively entrained from the disc. However, the absolute value of that (weakly spatially varying) FUV cross section depends sensitively on the grain properties in the disc itself. This is important for past and future modelling of external photoevaporative winds because it demonstrates that a single value for the FUV cross section can reasonably be assumed in the wind, even in multiple dimensions.  \\

In a companion paper we will use our models to study the predicted observational characteristics of the entrained dust, provide tests of our models and the means of interpreting observations of external photoevaporative winds.

\section*{Acknowledgements}
We thank Mark Hutchison for discussions and provision of data for the wind benchmark test comparison. We thank the referee for their positive review of our manuscript.

TJH acknowledges funding from a Royal Society Dorothy Hodgkin Fellowship, which also funds SP. TJH also acknowledges UKRI guaranteed funding for a Horizon Europe ERC consolidator grant (EP/Y024710/1). RPN acknowledges funding from STFC grants ST/X000931/1 and ST/T000341/1.

This research utilised Queen Mary University of London's Apocrita HPC facility, supported by QMUL Research-IT. http://doi.org/10.5281/zenodo.438045.

This work used the DiRAC Data Intensive service at Leicester, operated by the University of Leicester IT Services, which forms part of the STFC DiRAC HPC Facility (www.dirac.ac.uk). The equipment was funded by BEIS capital funding via STFC capital grants ST/K000373/1 and ST/R002363/1 and STFC DiRAC Operations grant ST/R001014/1. DiRAC is part of the National e-Infrastructure.

This work made significant use of Python 3 and the following packages: NumPy \cite{harris2020array}, Matplotlib \cite{Hunter:2007}, SciPy \cite{2020SciPy-NMeth} and Black \cite{BlackFormatter}. 
\section*{Data Availability}

The data associated with this paper will be made available upon reasonable request to the author.



\bibliographystyle{mnras}
\bibliography{2DPlanar} 

\begin{thebibliography}{}
\makeatletter
\relax
\def\mn@urlcharsother{\let\do\@makeother \do\$\do\&\do\#\do\^\do\_\do\%\do\~}
\def\mn@doi{\begingroup\mn@urlcharsother \@ifnextchar [ {\mn@doi@}
  {\mn@doi@[]}}
\def\mn@doi@[#1]#2{\def\@tempa{#1}\ifx\@tempa\@empty \href
  {http://dx.doi.org/#2} {doi:#2}\else \href {http://dx.doi.org/#2} {#1}\fi
  \endgroup}
\def\mn@eprint#1#2{\mn@eprint@#1:#2::\@nil}
\def\mn@eprint@arXiv#1{\href {http://arxiv.org/abs/#1} {{\tt arXiv:#1}}}
\def\mn@eprint@dblp#1{\href {http://dblp.uni-trier.de/rec/bibtex/#1.xml}
  {dblp:#1}}
\def\mn@eprint@#1:#2:#3:#4\@nil{\def\@tempa {#1}\def\@tempb {#2}\def\@tempc
  {#3}\ifx \@tempc \@empty \let \@tempc \@tempb \let \@tempb \@tempa \fi \ifx
  \@tempb \@empty \def\@tempb {arXiv}\fi \@ifundefined
  {mn@eprint@\@tempb}{\@tempb:\@tempc}{\expandafter \expandafter \csname
  mn@eprint@\@tempb\endcsname \expandafter{\@tempc}}}

\bibitem[\protect\citeauthoryear{{Adams}, {Hollenbach}, {Laughlin}  \&
  {Gorti}}{{Adams} et~al.}{2004}]{2004ApJ...611..360A}
{Adams} F.~C.,  {Hollenbach} D.,  {Laughlin} G.,   {Gorti} U.,  2004, \mn@doi
  [\apj] {10.1086/421989}, \href
  {https://ui.adsabs.harvard.edu/abs/2004ApJ...611..360A} {611, 360}

\bibitem[\protect\citeauthoryear{{Andrews} et~al.,}{{Andrews}
  et~al.}{2018}]{2018ApJ...869L..41A}
{Andrews} S.~M.,  et~al., 2018, \mn@doi [\apjl] {10.3847/2041-8213/aaf741},
  \href {https://ui.adsabs.harvard.edu/abs/2018ApJ...869L..41A} {869, L41}

\bibitem[\protect\citeauthoryear{{Aru} et~al.,}{{Aru}
  et~al.}{2024}]{2024arXiv240312604A}
{Aru} M.-L.,  et~al., 2024, \mn@doi [arXiv e-prints]
  {10.48550/arXiv.2403.12604}, \href
  {https://ui.adsabs.harvard.edu/abs/2024arXiv240312604A} {p. arXiv:2403.12604}

\bibitem[\protect\citeauthoryear{{Ballabio}, {Haworth}  \& {Henney}}{{Ballabio}
  et~al.}{2023}]{2023MNRAS.518.5563B}
{Ballabio} G.,  {Haworth} T.~J.,   {Henney} W.~J.,  2023, \mn@doi [\mnras]
  {10.1093/mnras/stac3467}, \href
  {https://ui.adsabs.harvard.edu/abs/2023MNRAS.518.5563B} {518, 5563}

\bibitem[\protect\citeauthoryear{{Ballering} et~al.,}{{Ballering}
  et~al.}{2023}]{2023ApJ...954..127B}
{Ballering} N.~P.,  et~al., 2023, \mn@doi [\apj] {10.3847/1538-4357/ace901},
  \href {https://ui.adsabs.harvard.edu/abs/2023ApJ...954..127B} {954, 127}

\bibitem[\protect\citeauthoryear{{Ballering}, {Cleeves}, {Boyden},
  {McCaughrean}, {Gross}  \& {Pearson}}{{Ballering}
  et~al.}{2024}]{2024arXiv241204356B}
{Ballering} N.~P.,  {Cleeves} L.~I.,  {Boyden} R.~D.,  {McCaughrean} M.~J.,
  {Gross} R.~E.,   {Pearson} S.~G.,  2024, \mn@doi [arXiv e-prints]
  {10.48550/arXiv.2412.04356}, \href
  {https://ui.adsabs.harvard.edu/abs/2024arXiv241204356B} {p. arXiv:2412.04356}

\bibitem[\protect\citeauthoryear{{Bally}, {O'Dell}  \& {McCaughrean}}{{Bally}
  et~al.}{2000}]{2000AJ....119.2919B}
{Bally} J.,  {O'Dell} C.~R.,   {McCaughrean} M.~J.,  2000, \mn@doi [\aj]
  {10.1086/301385}, \href
  {https://ui.adsabs.harvard.edu/abs/2000AJ....119.2919B} {119, 2919}

\bibitem[\protect\citeauthoryear{{Banzatti} et~al.,}{{Banzatti}
  et~al.}{2023}]{2023ApJ...957L..22B}
{Banzatti} A.,  et~al., 2023, \mn@doi [\apjl] {10.3847/2041-8213/acf5ec}, \href
  {https://ui.adsabs.harvard.edu/abs/2023ApJ...957L..22B} {957, L22}

\bibitem[\protect\citeauthoryear{{Bern{\'e}} et~al.,}{{Bern{\'e}}
  et~al.}{2024}]{2024Sci...383..988B}
{Bern{\'e}} O.,  et~al., 2024, \mn@doi [Science] {10.1126/science.adh2861},
  \href {https://ui.adsabs.harvard.edu/abs/2024Sci...383..988B} {383, 988}

\bibitem[\protect\citeauthoryear{{Birnstiel}, {Klahr}  \&
  {Ercolano}}{{Birnstiel} et~al.}{2012}]{2012A&A...539A.148B}
{Birnstiel} T.,  {Klahr} H.,   {Ercolano} B.,  2012, \mn@doi [\aap]
  {10.1051/0004-6361/201118136}, \href
  {https://ui.adsabs.harvard.edu/abs/2012A&A...539A.148B} {539, A148}

\bibitem[\protect\citeauthoryear{{Bisbas}, {Haworth}, {Barlow}, {Viti},
  {Harries}, {Bell}  \& {Yates}}{{Bisbas} et~al.}{2015}]{2015MNRAS.454.2828B}
{Bisbas} T.~G.,  {Haworth} T.~J.,  {Barlow} M.~J.,  {Viti} S.,  {Harries}
  T.~J.,  {Bell} T.,   {Yates} J.~A.,  2015, \mn@doi [\mnras]
  {10.1093/mnras/stv2156}, \href
  {https://ui.adsabs.harvard.edu/abs/2015MNRAS.454.2828B} {454, 2828}

\bibitem[\protect\citeauthoryear{{Booth} \& {Clarke}}{{Booth} \&
  {Clarke}}{2021}]{2021MNRAS.502.1569B}
{Booth} R.~A.,  {Clarke} C.~J.,  2021, \mn@doi [\mnras]
  {10.1093/mnras/stab090}, \href
  {https://ui.adsabs.harvard.edu/abs/2021MNRAS.502.1569B} {502, 1569}

\bibitem[\protect\citeauthoryear{{Burn}, {Emsenhuber}, {Weder}, {V{\"o}lkel},
  {Klahr}, {Birnstiel}, {Ercolano}  \& {Mordasini}}{{Burn}
  et~al.}{2022}]{2022A&A...666A..73B}
{Burn} R.,  {Emsenhuber} A.,  {Weder} J.,  {V{\"o}lkel} O.,  {Klahr} H.,
  {Birnstiel} T.,  {Ercolano} B.,   {Mordasini} C.,  2022, \mn@doi [\aap]
  {10.1051/0004-6361/202243262}, \href
  {https://ui.adsabs.harvard.edu/abs/2022A&A...666A..73B} {666, A73}

\bibitem[\protect\citeauthoryear{{Clarke} \& {Alexander}}{{Clarke} \&
  {Alexander}}{2016}]{2016MNRAS.460.3044C}
{Clarke} C.~J.,  {Alexander} R.~D.,  2016, \mn@doi [\mnras]
  {10.1093/mnras/stw1178}, \href
  {https://ui.adsabs.harvard.edu/abs/2016MNRAS.460.3044C} {460, 3044}

\bibitem[\protect\citeauthoryear{{Coleman}, {Mroueh}  \& {Haworth}}{{Coleman}
  et~al.}{2024}]{2024MNRAS.527.7588C}
{Coleman} G. A.~L.,  {Mroueh} J.~K.,   {Haworth} T.~J.,  2024, \mn@doi [\mnras]
  {10.1093/mnras/stad3692}, \href
  {https://ui.adsabs.harvard.edu/abs/2024MNRAS.527.7588C} {527, 7588}

\bibitem[\protect\citeauthoryear{{Draine} \& {Lee}}{{Draine} \&
  {Lee}}{1984}]{1984ApJ...285...89D}
{Draine} B.~T.,  {Lee} H.~M.,  1984, \mn@doi [\apj] {10.1086/162480}, \href
  {https://ui.adsabs.harvard.edu/abs/1984ApJ...285...89D} {285, 89}

\bibitem[\protect\citeauthoryear{{Dullemond} \& {Dominik}}{{Dullemond} \&
  {Dominik}}{2004}]{2004Dullemond&Dominik}
{Dullemond} C.~P.,  {Dominik} C.,  2004, \mn@doi [\aap]
  {10.1051/0004-6361:20040284}, \href
  {https://ui.adsabs.harvard.edu/abs/2004A&A...421.1075D} {421, 1075}

\bibitem[\protect\citeauthoryear{{Dullemond} et~al.,}{{Dullemond}
  et~al.}{2018}]{2018ApJ...869L..46D}
{Dullemond} C.~P.,  et~al., 2018, \mn@doi [\apjl] {10.3847/2041-8213/aaf742},
  \href {https://ui.adsabs.harvard.edu/abs/2018ApJ...869L..46D} {869, L46}

\bibitem[\protect\citeauthoryear{{Facchini}, {Clarke}  \& {Bisbas}}{{Facchini}
  et~al.}{2016}]{2016MNRAS.457.3593F}
{Facchini} S.,  {Clarke} C.~J.,   {Bisbas} T.~G.,  2016, \mn@doi [\mnras]
  {10.1093/mnras/stw240}, \href
  {https://ui.adsabs.harvard.edu/abs/2016MNRAS.457.3593F} {457, 3593}

\bibitem[\protect\citeauthoryear{{G{\'a}rate}, {Pinilla}, {Haworth}  \&
  {Facchini}}{{G{\'a}rate} et~al.}{2023a}]{2023arXiv231020214G}
{G{\'a}rate} M.,  {Pinilla} P.,  {Haworth} T.~J.,   {Facchini} S.,  2023a,
  \mn@doi [arXiv e-prints] {10.48550/arXiv.2310.20214}, \href
  {https://ui.adsabs.harvard.edu/abs/2023arXiv231020214G} {p. arXiv:2310.20214}

\bibitem[\protect\citeauthoryear{{G{\'a}rate} et~al.,}{{G{\'a}rate}
  et~al.}{2023b}]{2023A&A...679A..15G}
{G{\'a}rate} M.,  et~al., 2023b, \mn@doi [\aap] {10.1051/0004-6361/202244436},
  \href {https://ui.adsabs.harvard.edu/abs/2023A&A...679A..15G} {679, A15}

\bibitem[\protect\citeauthoryear{{Garc{\'\i}a-Arredondo}, {Henney}  \&
  {Arthur}}{{Garc{\'\i}a-Arredondo} et~al.}{2001}]{2001ApJ...561..830G}
{Garc{\'\i}a-Arredondo} F.,  {Henney} W.~J.,   {Arthur} S.~J.,  2001, \mn@doi
  [\apj] {10.1086/323367}, \href
  {https://ui.adsabs.harvard.edu/abs/2001ApJ...561..830G} {561, 830}

\bibitem[\protect\citeauthoryear{{Gasman} et~al.,}{{Gasman}
  et~al.}{2025}]{2025arXiv250104587G}
{Gasman} D.,  et~al., 2025, arXiv e-prints, \href
  {https://ui.adsabs.harvard.edu/abs/2025arXiv250104587G} {p. arXiv:2501.04587}

\bibitem[\protect\citeauthoryear{{Gorti} \& {Hollenbach}}{{Gorti} \&
  {Hollenbach}}{2009}]{2009ApJ...690.1539G}
{Gorti} U.,  {Hollenbach} D.,  2009, \mn@doi [\apj]
  {10.1088/0004-637X/690/2/1539}, \href
  {https://ui.adsabs.harvard.edu/abs/2009ApJ...690.1539G} {690, 1539}

\bibitem[\protect\citeauthoryear{{Harries}}{{Harries}}{2015}]{2015MNRAS.448.3156H}
{Harries} T.~J.,  2015, \mn@doi [\mnras] {10.1093/mnras/stv158}, \href
  {https://ui.adsabs.harvard.edu/abs/2015MNRAS.448.3156H} {448, 3156}

\bibitem[\protect\citeauthoryear{{Harries}, {Haworth}, {Acreman}, {Ali}  \&
  {Douglas}}{{Harries} et~al.}{2019}]{2019A&C....27...63H}
{Harries} T.~J.,  {Haworth} T.~J.,  {Acreman} D.,  {Ali} A.,   {Douglas} T.,
  2019, \mn@doi [Astronomy and Computing] {10.1016/j.ascom.2019.03.002}, \href
  {https://ui.adsabs.harvard.edu/abs/2019A&C....27...63H} {27, 63}

\bibitem[\protect\citeauthoryear{Harris et~al.,}{Harris
  et~al.}{2020}]{harris2020array}
Harris C.~R.,  et~al., 2020, \mn@doi [Nature] {10.1038/s41586-020-2649-2}, 585,
  357

\bibitem[\protect\citeauthoryear{{Haworth} \& {Clarke}}{{Haworth} \&
  {Clarke}}{2019}]{2019MNRAS.485.3895H}
{Haworth} T.~J.,  {Clarke} C.~J.,  2019, \mn@doi [\mnras]
  {10.1093/mnras/stz706}, \href
  {https://ui.adsabs.harvard.edu/abs/2019MNRAS.485.3895H} {485, 3895}

\bibitem[\protect\citeauthoryear{{Haworth}, {Boubert}, {Facchini}, {Bisbas}  \&
  {Clarke}}{{Haworth} et~al.}{2016}]{2016MNRAS.463.3616H}
{Haworth} T.~J.,  {Boubert} D.,  {Facchini} S.,  {Bisbas} T.~G.,   {Clarke}
  C.~J.,  2016, \mn@doi [\mnras] {10.1093/mnras/stw2280}, \href
  {https://ui.adsabs.harvard.edu/abs/2016MNRAS.463.3616H} {463, 3616}

\bibitem[\protect\citeauthoryear{{Haworth}, {Facchini}, {Clarke}  \&
  {Cleeves}}{{Haworth} et~al.}{2017}]{2017MNRAS.468L.108H}
{Haworth} T.~J.,  {Facchini} S.,  {Clarke} C.~J.,   {Cleeves} L.~I.,  2017,
  \mn@doi [\mnras] {10.1093/mnrasl/slx037}, \href
  {https://ui.adsabs.harvard.edu/abs/2017MNRAS.468L.108H} {468, L108}

\bibitem[\protect\citeauthoryear{{Haworth}, {Clarke}, {Rahman}, {Winter}  \&
  {Facchini}}{{Haworth} et~al.}{2018}]{2018MNRAS.481..452H}
{Haworth} T.~J.,  {Clarke} C.~J.,  {Rahman} W.,  {Winter} A.~J.,   {Facchini}
  S.,  2018, \mn@doi [\mnras] {10.1093/mnras/sty2323}, \href
  {https://ui.adsabs.harvard.edu/abs/2018MNRAS.481..452H} {481, 452}

\bibitem[\protect\citeauthoryear{{Haworth}, {Coleman}, {Qiao}, {Sellek}  \&
  {Askari}}{{Haworth} et~al.}{2023}]{2023MNRAS.526.4315H}
{Haworth} T.~J.,  {Coleman} G. A.~L.,  {Qiao} L.,  {Sellek} A.~D.,   {Askari}
  K.,  2023, \mn@doi [\mnras] {10.1093/mnras/stad3054}, \href
  {https://ui.adsabs.harvard.edu/abs/2023MNRAS.526.4315H} {526, 4315}

\bibitem[\protect\citeauthoryear{{Hayward}, {Houck}  \& {Miles}}{{Hayward}
  et~al.}{1994}]{1994ApJ...433..157H}
{Hayward} T.~L.,  {Houck} J.~R.,   {Miles} J.~W.,  1994, \mn@doi [\apj]
  {10.1086/174632}, \href
  {https://ui.adsabs.harvard.edu/abs/1994ApJ...433..157H} {433, 157}

\bibitem[\protect\citeauthoryear{{Hollenbach}, {Johnstone}, {Lizano}  \&
  {Shu}}{{Hollenbach} et~al.}{1994}]{1994ApJ...428..654H}
{Hollenbach} D.,  {Johnstone} D.,  {Lizano} S.,   {Shu} F.,  1994, \mn@doi
  [\apj] {10.1086/174276}, \href
  {https://ui.adsabs.harvard.edu/abs/1994ApJ...428..654H} {428, 654}

\bibitem[\protect\citeauthoryear{Hunter}{Hunter}{2007}]{Hunter:2007}
Hunter J.~D.,  2007, \mn@doi [Computing in Science \& Engineering]
  {10.1109/MCSE.2007.55}, 9, 90

\bibitem[\protect\citeauthoryear{{Hutchison} \& {Clarke}}{{Hutchison} \&
  {Clarke}}{2021}]{2021MNRAS.501.1127H}
{Hutchison} M.~A.,  {Clarke} C.~J.,  2021, \mn@doi [\mnras]
  {10.1093/mnras/staa3608}, \href
  {https://ui.adsabs.harvard.edu/abs/2021MNRAS.501.1127H} {501, 1127}

\bibitem[\protect\citeauthoryear{{Hutchison}, {Price}, {Laibe}  \&
  {Maddison}}{{Hutchison} et~al.}{2016a}]{2016MNRAS.461..742H}
{Hutchison} M.~A.,  {Price} D.~J.,  {Laibe} G.,   {Maddison} S.~T.,  2016a,
  \mn@doi [\mnras] {10.1093/mnras/stw1126}, \href
  {https://ui.adsabs.harvard.edu/abs/2016MNRAS.461..742H} {461, 742}

\bibitem[\protect\citeauthoryear{{Hutchison}, {Laibe}  \&
  {Maddison}}{{Hutchison} et~al.}{2016b}]{2016MNRAS.463.2725H}
{Hutchison} M.~A.,  {Laibe} G.,   {Maddison} S.~T.,  2016b, \mn@doi [\mnras]
  {10.1093/mnras/stw2191}, \href
  {https://ui.adsabs.harvard.edu/abs/2016MNRAS.463.2725H} {463, 2725}

\bibitem[\protect\citeauthoryear{{Jang}, {Waters}, {Kamp}  \&
  {Dullemond}}{{Jang} et~al.}{2024}]{2024A&A...687A.275J}
{Jang} H.,  {Waters} R.,  {Kamp} I.,   {Dullemond} C.~P.,  2024, \mn@doi [\aap]
  {10.1051/0004-6361/202348630}, \href
  {https://ui.adsabs.harvard.edu/abs/2024A&A...687A.275J} {687, A275}

\bibitem[\protect\citeauthoryear{{Johansen} \& {Lambrechts}}{{Johansen} \&
  {Lambrechts}}{2017}]{2017AREPS..45..359J}
{Johansen} A.,  {Lambrechts} M.,  2017, \mn@doi [Annual Review of Earth and
  Planetary Sciences] {10.1146/annurev-earth-063016-020226}, \href
  {https://ui.adsabs.harvard.edu/abs/2017AREPS..45..359J} {45, 359}

\bibitem[\protect\citeauthoryear{{Johnstone}, {Hollenbach}  \&
  {Bally}}{{Johnstone} et~al.}{1998}]{1998ApJ...499..758J}
{Johnstone} D.,  {Hollenbach} D.,   {Bally} J.,  1998, \mn@doi [\apj]
  {10.1086/305658}, \href
  {https://ui.adsabs.harvard.edu/abs/1998ApJ...499..758J} {499, 758}

\bibitem[\protect\citeauthoryear{{Kim}, {Clarke}, {Fang}  \& {Facchini}}{{Kim}
  et~al.}{2016}]{2016ApJ...826L..15K}
{Kim} J.~S.,  {Clarke} C.~J.,  {Fang} M.,   {Facchini} S.,  2016, \mn@doi
  [\apjl] {10.3847/2041-8205/826/1/L15}, \href
  {https://ui.adsabs.harvard.edu/abs/2016ApJ...826L..15K} {826, L15}

\bibitem[\protect\citeauthoryear{Krijt, Schwarz, Bergin  \& Ciesla}{Krijt
  et~al.}{2018}]{2018Krijt+}
Krijt S.,  Schwarz K.~R.,  Bergin E.~A.,   Ciesla F.~J.,  2018, \mn@doi [The
  Astrophysical Journal] {10.3847/1538-4357/aad69b}, 864, 78

\bibitem[\protect\citeauthoryear{{Lambrechts} \& {Johansen}}{{Lambrechts} \&
  {Johansen}}{2012}]{2012A&A...544A..32L}
{Lambrechts} M.,  {Johansen} A.,  2012, \mn@doi [\aap]
  {10.1051/0004-6361/201219127}, \href
  {https://ui.adsabs.harvard.edu/abs/2012A&A...544A..32L} {544, A32}

\bibitem[\protect\citeauthoryear{{Lambrechts}, {Johansen}  \&
  {Morbidelli}}{{Lambrechts} et~al.}{2014}]{2014A&A...572A..35L}
{Lambrechts} M.,  {Johansen} A.,   {Morbidelli} A.,  2014, \mn@doi [\aap]
  {10.1051/0004-6361/201423814}, \href
  {https://ui.adsabs.harvard.edu/abs/2014A&A...572A..35L} {572, A35}

\bibitem[\protect\citeauthoryear{Langa \& contributors~to Black}{Langa \&
  contributors~to Black}{}]{BlackFormatter}
Langa L.,  contributors~to Black, Black: The uncompromising Python code
  formatter, \url {https://black.readthedocs.io/en/stable/}

\bibitem[\protect\citeauthoryear{{Manara}, {Ansdell}, {Rosotti}, {Hughes},
  {Armitage}, {Lodato}  \& {Williams}}{{Manara}
  et~al.}{2023}]{2023ASPC..534..539M}
{Manara} C.~F.,  {Ansdell} M.,  {Rosotti} G.~P.,  {Hughes} A.~M.,  {Armitage}
  P.~J.,  {Lodato} G.,   {Williams} J.~P.,  2023, in {Inutsuka} S.,  {Aikawa}
  Y.,  {Muto} T.,  {Tomida} K.,   {Tamura} M.,  eds,  Astronomical Society of
  the Pacific Conference Series Vol. 534, Protostars and Planets VII. p.~539
  (\mn@eprint {arXiv} {2203.09930}), \mn@doi{10.48550/arXiv.2203.09930}

\bibitem[\protect\citeauthoryear{{Michel}, {van der Marel}  \&
  {Matthews}}{{Michel} et~al.}{2021}]{2021ApJ...921...72M}
{Michel} A.,  {van der Marel} N.,   {Matthews} B.~C.,  2021, \mn@doi [\apj]
  {10.3847/1538-4357/ac1bbb}, \href
  {https://ui.adsabs.harvard.edu/abs/2021ApJ...921...72M} {921, 72}

\bibitem[\protect\citeauthoryear{{Miotello}, {Robberto}, {Potenza}  \&
  {Ricci}}{{Miotello} et~al.}{2012}]{2012ApJ...757...78M}
{Miotello} A.,  {Robberto} M.,  {Potenza} M. A.~C.,   {Ricci} L.,  2012,
  \mn@doi [\apj] {10.1088/0004-637X/757/1/78}, \href
  {https://ui.adsabs.harvard.edu/abs/2012ApJ...757...78M} {757, 78}

\bibitem[\protect\citeauthoryear{{Nakatani}, {Hosokawa}, {Yoshida}, {Nomura}
  \& {Kuiper}}{{Nakatani} et~al.}{2018}]{2018ApJ...857...57N}
{Nakatani} R.,  {Hosokawa} T.,  {Yoshida} N.,  {Nomura} H.,   {Kuiper} R.,
  2018, \mn@doi [\apj] {10.3847/1538-4357/aab70b}, \href
  {https://ui.adsabs.harvard.edu/abs/2018ApJ...857...57N} {857, 57}

\bibitem[\protect\citeauthoryear{{Nayakshin}, {Cha}  \& {Hobbs}}{{Nayakshin}
  et~al.}{2009}]{2009MNRAS.397.1314N}
{Nayakshin} S.,  {Cha} S.-H.,   {Hobbs} A.,  2009, \mn@doi [\mnras]
  {10.1111/j.1365-2966.2009.15091.x}, \href
  {https://ui.adsabs.harvard.edu/abs/2009MNRAS.397.1314N} {397, 1314}

\bibitem[\protect\citeauthoryear{{O'Dell} \& {Wen}}{{O'Dell} \&
  {Wen}}{1994}]{1994ApJ...436..194O}
{O'Dell} C.~R.,  {Wen} Z.,  1994, \mn@doi [\apj] {10.1086/174892}, \href
  {https://ui.adsabs.harvard.edu/abs/1994ApJ...436..194O} {436, 194}

\bibitem[\protect\citeauthoryear{{O'Dell}, {Wen}  \& {Hu}}{{O'Dell}
  et~al.}{1993}]{1993ApJ...410..696O}
{O'Dell} C.~R.,  {Wen} Z.,   {Hu} X.,  1993, \mn@doi [\apj] {10.1086/172786},
  \href {https://ui.adsabs.harvard.edu/abs/1993ApJ...410..696O} {410, 696}

\bibitem[\protect\citeauthoryear{{{\"O}berg} et~al.,}{{{\"O}berg}
  et~al.}{2021}]{2021ApJS..257....1O}
{{\"O}berg} K.~I.,  et~al., 2021, \mn@doi [\apjs] {10.3847/1538-4365/ac1432},
  \href {https://ui.adsabs.harvard.edu/abs/2021ApJS..257....1O} {257, 1}

\bibitem[\protect\citeauthoryear{{Owen} \& {Altaf}}{{Owen} \&
  {Altaf}}{2021}]{2021MNRAS.508.2493O}
{Owen} J.~E.,  {Altaf} N.,  2021, \mn@doi [\mnras] {10.1093/mnras/stab2749},
  \href {https://ui.adsabs.harvard.edu/abs/2021MNRAS.508.2493O} {508, 2493}

\bibitem[\protect\citeauthoryear{{Owen}, {Clarke}  \& {Ercolano}}{{Owen}
  et~al.}{2012}]{2012MNRAS.422.1880O}
{Owen} J.~E.,  {Clarke} C.~J.,   {Ercolano} B.,  2012, \mn@doi [\mnras]
  {10.1111/j.1365-2966.2011.20337.x}, \href
  {https://ui.adsabs.harvard.edu/abs/2012MNRAS.422.1880O} {422, 1880}

\bibitem[\protect\citeauthoryear{{Pascucci}, {Cabrit}, {Edwards}, {Gorti},
  {Gressel}  \& {Suzuki}}{{Pascucci} et~al.}{2023}]{2023ASPC..534..567P}
{Pascucci} I.,  {Cabrit} S.,  {Edwards} S.,  {Gorti} U.,  {Gressel} O.,
  {Suzuki} T.~K.,  2023, in {Inutsuka} S.,  {Aikawa} Y.,  {Muto} T.,  {Tomida}
  K.,   {Tamura} M.,  eds,  Astronomical Society of the Pacific Conference
  Series Vol. 534, Protostars and Planets VII. p.~567 (\mn@eprint {arXiv}
  {2203.10068}), \mn@doi{10.48550/arXiv.2203.10068}

\bibitem[\protect\citeauthoryear{{Picogna}, {Ercolano}, {Owen}  \&
  {Weber}}{{Picogna} et~al.}{2019}]{2019MNRAS.487..691P}
{Picogna} G.,  {Ercolano} B.,  {Owen} J.~E.,   {Weber} M.~L.,  2019, \mn@doi
  [\mnras] {10.1093/mnras/stz1166}, \href
  {https://ui.adsabs.harvard.edu/abs/2019MNRAS.487..691P} {487, 691}

\bibitem[\protect\citeauthoryear{{Qiao}, {Coleman}  \& {Haworth}}{{Qiao}
  et~al.}{2023}]{2023MNRAS.522.1939Q}
{Qiao} L.,  {Coleman} G. A.~L.,   {Haworth} T.~J.,  2023, \mn@doi [\mnras]
  {10.1093/mnras/stad944}, \href
  {https://ui.adsabs.harvard.edu/abs/2023MNRAS.522.1939Q} {522, 1939}

\bibitem[\protect\citeauthoryear{{Ricci}, {Robberto}  \& {Soderblom}}{{Ricci}
  et~al.}{2008}]{2008AJ....136.2136R}
{Ricci} L.,  {Robberto} M.,   {Soderblom} D.~R.,  2008, \mn@doi [\aj]
  {10.1088/0004-6256/136/5/2136}, \href
  {https://ui.adsabs.harvard.edu/abs/2008AJ....136.2136R} {136, 2136}

\bibitem[\protect\citeauthoryear{{Richling} \& {Yorke}}{{Richling} \&
  {Yorke}}{2000}]{2000ApJ...539..258R}
{Richling} S.,  {Yorke} H.~W.,  2000, \mn@doi [\apj] {10.1086/309198}, \href
  {https://ui.adsabs.harvard.edu/abs/2000ApJ...539..258R} {539, 258}

\bibitem[\protect\citeauthoryear{Rohatgi}{Rohatgi}{}]{WebPlotDigitizer}
Rohatgi A., , WebPlotDigitizer, \url {https://automeris.io}

\bibitem[\protect\citeauthoryear{{Sellek}, {Booth}  \& {Clarke}}{{Sellek}
  et~al.}{2020}]{2020MNRAS.492.1279S}
{Sellek} A.~D.,  {Booth} R.~A.,   {Clarke} C.~J.,  2020, \mn@doi [\mnras]
  {10.1093/mnras/stz3528}, \href
  {https://ui.adsabs.harvard.edu/abs/2020MNRAS.492.1279S} {492, 1279}

\bibitem[\protect\citeauthoryear{{Tabone}, {Rosotti}, {Lodato}, {Armitage},
  {Cridland}  \& {van Dishoeck}}{{Tabone} et~al.}{2022a}]{2022MNRAS.512L..74T}
{Tabone} B.,  {Rosotti} G.~P.,  {Lodato} G.,  {Armitage} P.~J.,  {Cridland}
  A.~J.,   {van Dishoeck} E.~F.,  2022a, \mn@doi [\mnras]
  {10.1093/mnrasl/slab124}, \href
  {https://ui.adsabs.harvard.edu/abs/2022MNRAS.512L..74T} {512, L74}

\bibitem[\protect\citeauthoryear{{Tabone}, {Rosotti}, {Cridland}, {Armitage}
  \& {Lodato}}{{Tabone} et~al.}{2022b}]{2022MNRAS.512.2290T}
{Tabone} B.,  {Rosotti} G.~P.,  {Cridland} A.~J.,  {Armitage} P.~J.,   {Lodato}
  G.,  2022b, \mn@doi [\mnras] {10.1093/mnras/stab3442}, \href
  {https://ui.adsabs.harvard.edu/abs/2022MNRAS.512.2290T} {512, 2290}

\bibitem[\protect\citeauthoryear{{Takeuchi}, {Clarke}  \& {Lin}}{{Takeuchi}
  et~al.}{2005}]{2005Takeuchi+}
{Takeuchi} T.,  {Clarke} C.~J.,   {Lin} D.~N.~C.,  2005, \mn@doi [\apj]
  {10.1086/430393}, \href
  {https://ui.adsabs.harvard.edu/abs/2005ApJ...627..286T} {627, 286}

\bibitem[\protect\citeauthoryear{{Testi} et~al.,}{{Testi}
  et~al.}{2014}]{2014prpl.conf..339T}
{Testi} L.,  et~al., 2014, in {Beuther} H.,  {Klessen} R.~S.,  {Dullemond}
  C.~P.,   {Henning} T.,  eds, Protostars and Planets VI. pp 339--361
  (\mn@eprint {arXiv} {1402.1354}),
  \mn@doi{10.2458/azu_uapress_9780816531240-ch015}

\bibitem[\protect\citeauthoryear{{Throop} \& {Bally}}{{Throop} \&
  {Bally}}{2005}]{2005Throop&Bally}
{Throop} H.~B.,  {Bally} J.,  2005, \mn@doi [\apjl] {10.1086/430272}, \href
  {https://ui.adsabs.harvard.edu/abs/2005ApJ...623L.149T} {623, L149}

\bibitem[\protect\citeauthoryear{Virtanen et~al.,}{Virtanen
  et~al.}{2020}]{2020SciPy-NMeth}
Virtanen P.,  et~al., 2020, \mn@doi [Nature Methods]
  {10.1038/s41592-019-0686-2}, \href {https://rdcu.be/b08Wh} {17, 261}

\bibitem[\protect\citeauthoryear{{Wang}, {Bai}  \& {Goodman}}{{Wang}
  et~al.}{2019}]{2019ApJ...874...90W}
{Wang} L.,  {Bai} X.-N.,   {Goodman} J.,  2019, \mn@doi [\apj]
  {10.3847/1538-4357/ab06fd}, \href
  {https://ui.adsabs.harvard.edu/abs/2019ApJ...874...90W} {874, 90}

\bibitem[\protect\citeauthoryear{{Weidenschilling}}{{Weidenschilling}}{1977}]{1977MNRAS.180...57W}
{Weidenschilling} S.~J.,  1977, \mn@doi [\mnras] {10.1093/mnras/180.2.57},
  \href {https://ui.adsabs.harvard.edu/abs/1977MNRAS.180...57W} {180, 57}

\bibitem[\protect\citeauthoryear{{Whipple}}{{Whipple}}{1972}]{1972fpp..conf..211W}
{Whipple} F.~L.,  1972, in {Elvius} A.,  ed., From Plasma to Planet. p.~211

\bibitem[\protect\citeauthoryear{{Winter} \& {Haworth}}{{Winter} \&
  {Haworth}}{2022}]{2022EPJP..137.1132W}
{Winter} A.~J.,  {Haworth} T.~J.,  2022, \mn@doi [European Physical Journal
  Plus] {10.1140/epjp/s13360-022-03314-1}, \href
  {https://ui.adsabs.harvard.edu/abs/2022EPJP..137.1132W} {137, 1132}

\bibitem[\protect\citeauthoryear{{van Terwisga} \& {Hacar}}{{van Terwisga} \&
  {Hacar}}{2023}]{2023A&A...673L...2V}
{van Terwisga} S.~E.,  {Hacar} A.,  2023, \mn@doi [\aap]
  {10.1051/0004-6361/202346135}, \href
  {https://ui.adsabs.harvard.edu/abs/2023A&A...673L...2V} {673, L2}

\makeatother
\end{thebibliography}

\appendix

\section{Particle solver methodology} 
\label{appendix:method}

\subsection{Cylindrical Equations of Motion}
The first benchmark for this code tested the gravity - ensuring stable, non-precessing orbits were produced over long periods of time.
We start by deriving the gravitational and fictitious forces in cylindrical coordinates, first defining the radial $\hat{R}$, angular ${\hat{\phi}}$ and vertical $\hat{z}$ positions and their respective velocities ($\frac{\partial x_i}{\partial t} \equiv \dot{\phantom{x}}$).

The basic unit vector properties are as follows, defined with their usual Cartesian counterparts $\hat{i}$, $\hat{j}$ and $\hat{k}$:
\begin{align}
    \hat{R} &= \cos{\phi} \hat{i} + \sin{\phi} \hat{j}\\
    \hat{\phi} &= -\sin{\phi} \hat{i} + \cos{\phi} \hat{j}\\
    \hat{z} &= \hat{k}\\
    \dot{\hat{R}} &= -\dot{\phi}\sin{\phi} \hat{i} + \dot{\phi}\cos{\phi} \hat{j} = \dot{\phi}\hat{\phi}\\
    \dot{\hat{\phi}} &= -\dot{\phi}\cos{\phi} \hat{i} - \dot{\phi}\sin{\phi} \hat{j} = -\dot{\phi}\hat{R}\\
    \dot{\hat{z}} &= 0
\end{align}
We then have the following base position vector for this system
\begin{align}
    \vec{r} &= R \hat{R} + z \hat{z},
\end{align}
velocity vector
\begin{align}
\vec{v} \equiv \dot{\vec{r}} &= \frac{d}{dt}(R \hat{R} + z \hat{z})\\
    &= \dot{R} \hat{R} + R \dot{\hat{R}} + z\dot{\hat{z}} + \dot{z}\hat{z} \nonumber\\
    &= \dot{R}\hat{R} + R\dot{\phi}\hat{\phi} + \dot{z}\hat{z}
\end{align}
and finally the acceleration vector
\begin{align}
    \vec{a} \equiv \dot{\vec{v}} &= \ddot{\vec{r}}\\
    &= \ddot{R}\hat{R} + \dot{R}\dot{\hat{R}} + \dot{R}\dot{\phi}\hat{\phi} + R\ddot{\phi}\hat{\phi} + R\dot{\phi}\dot{\hat{\phi}} + \ddot{z}\hat{z} + 0 \nonumber\\
    &= (\ddot{R} - R\dot{\phi}^2)\hat{R} + (2\dot{R}\dot{\phi} + R\ddot{\phi})\hat{\phi} + \ddot{z}\hat{z}.
\end{align}
With our basic equations of motion defined, we can now consider physical forces.

\subsection{Gravity}
First considering gravity, the fictitious forces are clearly evident from the acceleration term: $R\dot{\phi}^2$ is the centrifugal force (angular momentum) and $2\dot{R}\dot{\phi}$ is the Coriolis force. Gravity only acts towards an object (so in $\vec{r}$), so will only have components of $R$ and $z$
\begin{align}
    \vec{a}_g &= -\frac{GM_*}{|\vec{r}|^2} \hat{r} = -\frac{GM_*}{(R^2 + z^2)^{3/2}} \langle R, 0, z\rangle.
\end{align}
We can then set this equal to the $\ddot{\vec{r}}$ we derived earlier:
\begin{align}
    \ddot{R} - R\dot{\phi}^2 &= -\frac{GM_*R}{(R^2 + z^2)^{3/2}}\\
    2\dot{R}\dot{\phi} + R\ddot{\phi} &= 0 \label{eqn:ang_mom}\\
    \ddot{z} &= -\frac{GM_*z}{(R^2 + z^2)^{3/2}}
\end{align}
From equation \ref{eqn:ang_mom}, we get conservation of angular momentum
\begin{align}
    2\dot{R}\dot{\phi} + R\ddot{\phi} = \frac{d}{dt}(R^2\dot{\phi}) = 0\\
    R^2\dot{\phi} \equiv h,
\end{align}
where $h$ is defined to be the angular momentum and is constant.
We can now rewrite the radial acceleration using this constant:
\begin{align}
    \dot{\phi} &= \frac{h}{R^2}\\
    \ddot{R} &= -\frac{GM_*R}{(R^2 + z^2)^{3/2}} + \frac{h^2}{R^3},
\end{align}
giving us the final set of gravity equations:
\begin{align}
    \left\langle \ddot{R}, R\ddot{\phi}, \ddot{z} \right\rangle &= \left\langle -\frac{GM_*R}{(R^2 + z^2)^{3/2}} + \frac{h^2}{R^3}, -2\dot{R}\dot{\phi},  -\frac{GM_*z}{(R^2 + z^2)^{3/2}} \right\rangle.
\end{align}
By convention, we define the Keplerian angular velocity as one that produces a circular orbit in the midplane, or equivalently as the angular velocity at which there is no radial acceleration:
\begin{align}
    \dot{\phi}_{\mathrm{kep}} = \left(\frac{GM_*}{R^3}\right)^{1/2}.
\end{align}
We then look at the second conserved quantity: energy. We can define this with 
\begin{align}
    \frac{dE}{dt} &= 0.
\end{align}
The total energy budget is converted between kinetic and potential energy as a particle orbits:
\begin{align}
    E &= K + U = \frac{1}{2}m_d|\vec{v}|^2 - \frac{GM_*m_d}{|\vec{r}|}.
\end{align}
(Note that $|\vec{v}| = \sqrt{\dot{R}^2 + (R\dot{\phi})^2 + \dot{z}^2}$).
Here we can see that if the total energy is negative, the orbit will be bound, and if it is zero or greater, the particle will escape. Finally, we can use this to derive the escape speed:
\begin{align}
    v_{\mathrm{esc}} &= \left({\frac{2GM_*}{\sqrt{R^2 + z^2}}}\right)^{1/2}.
\end{align}

 In circular orbits, the gravitational and centrifugal forces cancel out and any small floating point arithmetic errors tend to be dampened as the Coriolis force adjusts the particle's angular velocity. For eccentric orbits, the gravitational and centrifugal forces do not cancel, but their sum averages to zero over a single orbit.


Various orbits were tested of different distances from 1\,AU to 1000\,AU and eccentricities from 0 to 0.9; all over the course of 10,000 orbits. At 100\,AU, this represents over 10\,Myrs of time, which exceeds the typical lifetimes of proto-planetary discs being irradiated. We also tested particles starting at the escape velocity of the system and faster, finding they produced the appropriate parabolic and hyperbolic orbits. All these orbits were found to be stable, non-precessing and conserve energy and angular momentum to within 0.25\% error overall. 

\subsection{Aerodynamic Drag}
The next step was to benchmark the aerodynamic drag force, using the drag equations from \cite{1977MNRAS.180...57W}. For the drag force, the acceleration can act in all directions, since it opposes particle motion. First, we look at the Epstein drag regime:
\begin{align}
    \vec{F}_D &= \frac{4\pi}{3} \rho s^2 \bar{v}\Delta\vec{v} \label{eqn:epstein}\\
    \vec{a}_D &= \frac{\rho}{s\rho_s} \bar{v} \Delta\vec{v}.
\end{align}
The grain radius and density are denoted by $s$ and $\rho_s$, respectively, while the gas density is represented without a subscript $\rho$.
$\bar{v}$ is the mean thermal velocity of the gas and $\Delta\vec{v}$ is the vector velocity difference between the gas and dust velocities.
Throughout this paper, we assume spherical dust grains with uniform density, to get a dust mass of $m_d = \nicefrac{4\pi}{3} \rho_s s^3$. Throughout this paper, we also use grain `radius' and `size' interchangeably to mean the same thing.

The $\phi$ component of acceleration is now no longer zero, and splitting the vector equation component-wise gives the three equations of motion for drag:
\begin{align}
    \ddot{R} &= \frac{\rho}{s\rho_s} \Delta v_R\bar{v}\\
    R\ddot{\phi} + 2\dot{R}\dot{\phi} &= \frac{\rho}{s\rho_s} \Delta v_\phi\bar{v} \nonumber\\
    R\ddot{\phi} &= \frac{\rho}{s\rho_s} \Delta v_\phi\bar{v} - 2\dot{R}\dot{\phi}\\
    \ddot{z} &= \frac{\rho}{s\rho_s} \Delta v_z\bar{v}.
\end{align}
Combining these now gives us the full set of equations to numerically integrate:
\begin{align}
    \ddot{R} &= -\frac{GM_*R}{(R^2 + z^2)^{3/2}} + \frac{h^2}{R^3} + \frac{\rho}{s\rho_s} \Delta v_R\bar{v}\\
    R\ddot{\phi} &= \frac{\rho}{s\rho_s} \Delta v_\phi\bar{v} - 2\dot{R}\frac{h}{R^2}\\
    \ddot{z} &= -\frac{GM_*z}{(R^2 + z^2)^{3/2}} + \frac{\rho}{s\rho_s} \Delta v_z\bar{v}.
\end{align}
The other drag regime we will consider is Stokes' drag:
\begin{align}
    \vec{F}_D &= C_D\pi \rho s^2 \frac{|\Delta\vec{v}| \Delta\vec{v}}{2}.\label{eqn:stokes}\\
    \vec{a}_D &= \frac{3C_D\rho |\Delta\vec{v}| \Delta\vec{v}}{8s\rho_s}.
\end{align}
Similarly, combined equations of motion with Stokes' drag can be derived, giving:
\begin{align}
    \ddot{R} &= -\frac{GM_*R}{(R^2 + z^2)^{3/2}} + \frac{h^2}{R^3} + \frac{3 C_D\rho}{8 s\rho_s} \Delta v_R|\Delta\vec{v}|\\
    R\ddot{\phi} &= \frac{3 C_D\rho}{8 s\rho_s} \Delta v_\phi|\Delta\vec{v}| - 2\dot{R}\frac{h}{R^2}\\
    \ddot{z} &= -\frac{GM_*z}{(R^2 + z^2)^{3/2}} + \frac{3 C_D\rho}{8 s\rho_s} \Delta v_z|\Delta\vec{v}|.
\end{align}

With our assumption that the dust is perfectly spherical, we can use the drag coefficient $C_D$ relation from \cite{1977MNRAS.180...57W}, which can be written in terms of the Reynolds number $Re$ and dynamic viscosity $\eta$ of the gas: 
\begin{align}
    C_D &= \begin{cases}24Re^{-1} \ \ \mathrm{for} \ Re \leq 1\\
    24Re^{-0.6} \ \ \mathrm{for} \ 1 < Re \leq 784.508...\\
    0.44 \ \ \mathrm{for} \ Re > 784.508... \ .\end{cases}
\end{align}
The Reynolds number is defined as follows,
\begin{align}
    Re &= \frac{2s\rho |\Delta\vec{v}|}{\eta}\\
    \eta &= \frac{1}{2}\rho\lambda \bar{v}\label{eqn:eta}\\
    \lambda &= \frac{1}{\sigma n}\label{eqn:mfp}\\
    n &= \frac{\rho}{\mu m_H}\label{eqn:num_dens},
\end{align}
where $c_s$ is the sound speed, $\lambda$ is the mean free path of the dust in the gas, $\eta$ is the dynamical viscosity (quoted from \cite{1972fpp..conf..211W}), $\sigma$ is the collisional cross section, $n$ is the number density of the gas and $\mu$ is the mean molecular mass of the gas. For all our models, we set a collisional cross section of $\sigma = 3.85\times10^{-15} \mathrm{cm}^2$. With our assumption of an ideal gas, the viscosity will be independent of density. 

Finally, the $Re=784.508...$ boundary comes from setting the last two forces equal and finding which Reynolds number matches them:
\begin{align}
    0.44 &= 24Re^{-0.6}\\
    Re &= \left(\frac{24}{0.44}\right)^{1/0.6} = 784.508... \ .
\end{align}
The $Re=1$ boundary can be similarly derived.

The analytic disc model selected for the benchmark is the same used by \cite{1977MNRAS.180...57W}. It describes an ideal gas with surface density and temperature power law distributions:
\begin{align}
    \Sigma(R) &= \Sigma_0 \left(R/R_0\right)^{-a+1}\\
    T(R) &= T_0 \left(R/R_0\right)^{-m}. \label{eqn:temperature}
\end{align}
For our tests, we used the same parameters as \cite{1977MNRAS.180...57W}: $m=1$ and $a=2$ (which is equivalent to an adiabatic disc with $\gamma$=3/2), $T_0=600$K and $\Sigma_0 = 1000 \mathrm{g/cm^2}$. The density was then calculated using the scale height of the disc
\begin{align}
    \rho(R) &= \frac{\Sigma_0}{\sqrt{2\pi H(R)}}\\
    H(R) &= \frac{c_s}{v_{k}} \label{eqn:scale_height} \\
    c_s(R) &= \sqrt{\frac{k_B T(R)}{\mu(R)}}
\end{align}
where we choose a constant $\mu(R) = \mu = 2.25$ for the entire midplane (an un-ionised hydrogen and helium gas). The choice of $\mu$ and $\sigma$ is not specified in the original \cite{1977MNRAS.180...57W} paper, so these values were instead reverse-engineered. The molecular mass $\mu$ was constrained by computing the drift rate in the ``perturbed Keplerian'' case, using the equations from \cite{1977MNRAS.180...57W}:
\begin{align}
    \frac{dr}{dt} &= \frac{r}{t_e}\frac{\Delta g}{g}    
\end{align}
and the stopping time \ref{eqn:stop_time} in the $Re > 784.508...$ regime and fitting it to the $10^5$\,cm sized dust (equally could have been fit to the $10^6$\,cm dust or any points on the graph squarely in the Stokes $Re > 784.508...$ regime). The gas collisional cross section $\sigma$ was fit to the transition points between the Epstein regime and the Stokes $Re < 1$ regime for the $0.1$\,cm to $1000$\,cm sized dust. Combining the equations for the mean free path Eq. \ref{eqn:mfp} and number density Eq. \ref{eqn:num_dens}, and remembering from \cite{1977MNRAS.180...57W} that the two regimes meet at $\nicefrac{\lambda}{s}=\nicefrac{4}{9}$, we can get an equation solving for $\sigma$:
\begin{equation}
    \sigma = \frac{\mu m_H}{\nicefrac{4}{9}s \rho}
\end{equation}
where we use the density computed at the transition radius (which depends on $\mu$, but is independent of $\sigma$).

Assuming hydrostatic equilibrium, no convection or turbulence and radially decreasing temperature, density and pressure at the midplane, the gas velocity is derived to be sub-Keplerian - being held in orbit by the pressure gradient:
\begin{align}
\frac{\partial\rho}{\partial t} + \nabla\cdot(\rho\vec{v}) &= 0 \label{eqn:hydro1}\\
\rho \frac{\partial \vec{v}}{\partial t} + \rho (\vec{v}\cdot\nabla)\vec{v} &= -\nabla P + \rho\vec{g}.
\end{align}
Applying our hydrostatic equilibrium condition $\vec{v}=0$ and $\frac{\partial}{\partial t} = 0$, this simplifies into a single equation (Eq. \ref{eqn:hydro1} trivially simplifies to zero):
\begin{align}
    0 &= -\nabla P + \rho\vec{g}.
\end{align}
We then split this component-wise:
\begin{align}
    0 & =-\frac{\partial P}{\partial R} + \rho g_R\\
    0 &= -\frac{1}{R}\frac{\partial P}{\partial \phi}\\
    0 &= -\frac{\partial P}{\partial z} + \rho g_z.
\end{align}
From this we see that there can't be any pressure variations in $\phi$. Plugging in the values for gravity, we recover the balanced force equations in $R$ and $z$:
\begin{align}
    0 & =-\frac{\partial P}{\partial R} - \frac{\rho GM_*R}{\left(R^2+z^2\right)^{3/2}} + \rho R\omega^2_{\mathrm{gas}}\\
    0 &= -\frac{1}{R}\frac{\partial P}{\partial \phi}\\
    0 &= -\frac{\partial P}{\partial z} - \frac{\rho GM_*z}{\left(R^2+z^2\right)^{3/2}}.
\end{align}
Finally, this gives us the gas velocity (by substituting $v_{\mathrm{gas}} = R\omega_{\mathrm{gas}}$):
\begin{align}
\frac{v^2_{\mathrm{gas}}}{R} &= \frac{GM_*R}{\left(R^2 + z^2\right)^{3/2}} + \frac{1}{\rho}\frac{\partial P}{\partial R}.
\end{align}
Here, the gas velocity simplifies in the midplane:
\begin{align}
\frac{v^2_\mathrm{{gas}}}{R} &= \frac{v^2_\mathrm{{kep}}}{R} + \frac{1}{\rho}\frac{\partial P}{\partial R}.
\end{align}
We also recover the vertical density structure of the gas disc (using $\frac{\rho GM_*z}{\left(R^2+z^2\right)^{3/2}} \sim \rho\Omega^2_K z$ for $R \ll z$, where $\Omega_K=GM_*/R^3$ and $\frac{\partial P}{\partial z} = \frac{\partial P}{\partial \rho}\frac{\partial \rho}{\partial z}$ with $c^2_s = \frac{\partial P}{\partial \rho}$ for an ideal gas):
\begin{align}
    \frac{\partial P}{\partial z} &= - \frac{\rho GM_*z}{\left(R^2+z^2\right)^{3/2}}\\
    c^2_s \frac{\partial \rho}{\partial z} &\sim -\rho\Omega^2_K z\\
    \rho(z) &= \rho(z=0) \exp{\left(-\frac{z^2\Omega_K^2}{2 c^2_s}\right)}\\ 
    \rho(R,z) &= \rho(R)\exp{\left(-\frac{z^2}{2H^2} \right)}. \label{eqn:density}
\end{align}
(Using the scale height $H = \frac{c_s}{\Omega_K}$).
Finally, assuming an ideal gas and no self-gravity for the disc, the pressure gradient will be independent of the disc mass [\cite{1977MNRAS.180...57W}].  With the disc fully defined, we can derive an analytic formula for the pressure gradient in terms of our disc parameters. For an ideal gas, we can define the pressure as $P = \frac{k_B}{\mu m_H} \rho(R)T(R)$. Then, 
\begin{align}
    \frac{\partial P}{\partial R} &= \frac{k_B}{\mu m_H} \left( \rho(R) \frac{\partial T}{\partial R} + T(R) \frac{\partial \rho}{\partial R}\right).
\end{align}
We then take our definitions of temperature \ref{eqn:temperature} and density \ref{eqn:density} and differentiate:
\begin{align}
    \frac{\partial T}{\partial R} &= -\frac{mT_0}{R}T(R)\\
    \frac{\partial H}{\partial R} &= \frac{3 H(R)}{2R} + \frac{H(R)}{T(R)}\frac{\partial T}{\partial R}\\
    &= H(R)\left(\frac{3}{2R} + \frac{1}{T(R)}\frac{\partial T}{\partial R}\right)\\
    \frac{\partial \rho}{\partial R} &= \rho(R) \left( \frac{-a-1}{R} + \frac{z^2}{H(R)^3}\frac{\partial H}{\partial R}\right).
\end{align}
Where the latter equation simplifies at the midplane ($z=0$). 

The resulting difference between the Keplerian and gas velocities causes dust in the disc to experience a drag force, proportional to the difference of their velocities.
The next step is implementing the drag forces the dust will feel. There are two relevant drag regimes to consider here: Epstein drag and Stokes' drag (recall \ref{eqn:epstein} and \ref{eqn:stokes}).

The Epstein regime occurs when a particle moves relatively slowly through a gas, and is smaller than the gas mean free path: $v \ll \bar{v}$ and $\lambda > s$. The two regimes meet at $\sfrac{\lambda}{s} = \sfrac{4}{9}$. 
Very small and very large particles will feel very little drag force overall, and will slowly drift inwards. In between, the dust feels a relatively stronger drag force and will be dragged inwards at a much faster rate. 

We ran the code, simulating dust particles of varying sizes.
In Weidenschilling's benchmark, three distinct regimes are considered: ``small'' dust, ``intermediate'' dust and ``large'' dust. The ``small'' dust regime occurs when dust stays squarely in the Epstein regime during its evolution. Here the dust rapidly reaches a terminal velocity equal to the local gas velocity. The ``large'' dust regime occurs when the dust experiences only high Reynolds number Stokes' drag. Here the dust never settles into a terminal velocity - instead spiralling inwards and constantly oscillating in velocity. Here Weidenschilling considers the radial drift rate rather than the terminal velocity. Finally, in the middle we have the ``intermediate'' dust case, which behaves as a mix of the two regimes, but does not have a simple analytic solution for its drift rate.

The results of these simulations are shown in Figure \ref{fig:weidenschilling}. 
To a high degree of accuracy, the same lifetimes were recovered as were found by \cite{1977MNRAS.180...57W}, differing by less than 3\% on average, with a maximum discrepancy of less than 15\%. This was compared against values extracted from the original plot, using the WebPlotDigitizer tool \cite{WebPlotDigitizer}. This gave us confidence our code was calculating aerodynamic drag correctly across the range of dust sizes and gas densities of interest. 

\subsection{A Tale of Two Frogs}
For this code, a 2nd-order leapfrog-like integrator was implemented. It combines a traditional leapfrog integrator, while also adding an additional intermediate step to keep velocity and position synchronised when calculating acceleration, to avoid the fictitious forces causing numerical instability. (This arised due to the fictitious forces being functions of velocity, and not just position, which the leapfrog integrator is optimised to solve).
\begin{align}
    x_{i+\frac{1}{2}} &= x_i + v_i \frac{1}{2} \Delta t\\
    v_{i+\frac{1}{2}} &= v_i + a(x_i, v_i) \frac{1}{2} \Delta t\\
    v_{i+1} &= v_i + a(x_{i+\frac{1}{2}}, v_{i+\frac{1}{2}}) \Delta t\\
    x_{i+1} &= x_{i+\frac{1}{2}} + v_{i+1} \frac{1}{2} \Delta t.
\end{align}
While this reduced instabilities, it unfortunately increased the maximum error bound of the method (proportional to $\Delta t^2$ instead of $\Delta t^3$). This was mitigated by choosing smaller timesteps to ensure our simulations returned the accuracy we desired.
The error bounds are derived below:
We define the actual solution as $x(t)$ and the leapfrog solution as $\tilde{x}(t)$. We then can define the error of our method as follows:
\begin{align}
    \Delta e(t+\Delta t) &= |x(t+\Delta t) - \tilde{x}(t+\Delta t)|.
\end{align}
We can expand both solutions as Taylor polynomials, which will cancel in the above error function, leaving behind the highest order term which contributes to the error.
\begin{align}
    x(t+\Delta t) &= x(t) + \dot{x}(t)\Delta t + \frac{1}{2}\ddot{x}(t) \Delta t^2 + ... \\
    x(t+\Delta t) &= x(t) + v(t)\Delta t + \frac{1}{2}a(t) \Delta t^2 + ...
\end{align}
where we can replace the first and second time derivatives with velocity and acceleration. We then do much of the same for the leapfrog solution:
\begin{align}
    \tilde{x}(t + \Delta t) &= \tilde{x}(t+\frac{1}{2}\Delta t) + \tilde{v}(t+\Delta t) \frac{1}{2} \Delta t\\
    &= \tilde{x}(t) + \tilde{v}(t) \frac{1}{2}\Delta t + (\tilde{v}(t) + a(t+\frac{1}{2}\Delta t) \Delta t) \frac{1}{2} \Delta t\\
    \tilde{x}(t + \Delta t) &= \tilde{x}(t) + \tilde{v}(t)\Delta t + \frac{1}{2}a(t+\frac{1}{2}\Delta t)\Delta t^2
\end{align}
Plugging these back into the error equation, to first order it simplifies to:
\begin{align}
    \Delta e(t+\Delta t) &\approx \frac{1}{2}\Delta t^2 |a(t) - a(t+\frac{1}{2}\Delta t)| \label{eqn:full_error}\\
    \Delta e(t+\Delta t) &\thicksim \mathcal{O}(\Delta t^2).
\end{align}
This was also numerically verified in a few benchmark cases, and generally, reducing the timestep size reduced the error following this relation. 


The timesteps were chosen such they were always less than both the aerodynamic ($\tau_s$) and dynamical stopping times ($\tau_d$). These are defined as follows:
\begin{align}
    \tau_s &= \begin{cases}
    \frac{|\vec{v}|}{|\vec{a}_D|} \label{eqn:stop_time} \ \ \mathrm{for} \ |\vec{a}_D|\neq 0\\
    1.0 \ \ \mathrm{for} \ |\vec{a}_D|=0\\
    \end{cases}\\
    \tau_d &= \begin{cases}
    K_g\frac{|\vec{r}|}{|\vec{v}|} = K_g\frac{\sqrt{R^2 + z^2}}{\sqrt{\dot{R}^2 + (R\dot{\phi})^2 + \dot{z}^2}} \ \ \mathrm{for} 
 \ |\vec{v}| \geq 55.0 \label{eqn:dyn_time}\\
    K_g\sqrt{\frac{R^3}{GM}} \ \ \mathrm{for} 
 \ |\vec{v}|<55.0
 \end{cases}
\end{align}
where we set $K_g=\frac{\pi}{4000\sqrt{2}}$, as we found this to produce orbits to our desired accuracy.

Also, in the code, all the angular positions, velocities and accelerations are handled in radians directly. (e.g. the Weidenschilling gas velocity becomes $\dot{\phi}^2_g = \dot{\phi}^2_{kep} + \frac{\Delta g}{\rho}$).

\subsection{Maximum Grain Size Solver} \label{appendix:favourite_section}
\begin{figure}
    \centering
    \includegraphics[width=1\columnwidth]{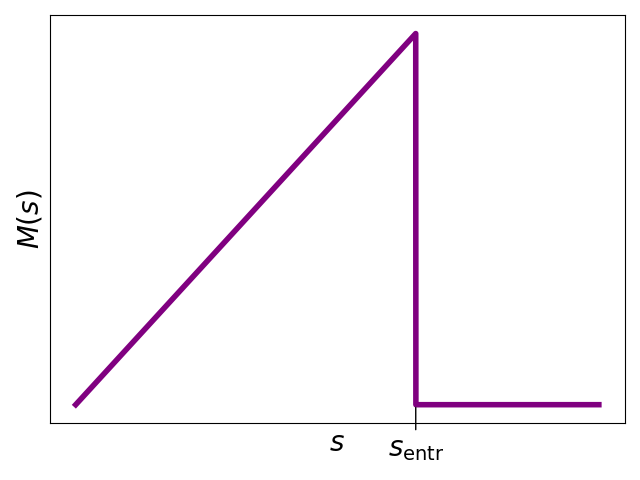}
    \caption{The maximum grain size function for a generic wind (equation \ref{eqn:max_dust}). Shows a single maximum value, when the grain size is equal to the maximum entrainable grain size ($s_{\textrm{entr}}$) of the wind.}
    \label{fig:dustsolver}
\end{figure}

A key quantity we wish to determine in photoevaporative disc models is the maximum entrained dust size.
In 1D flows the maximum grain size entrained from a starting position is assumed to be unimodal (i.e. there is only one maximum grain size and all grains larger than it are not entrained). A minimum requirement of this is that the flow cannot have any negative velocities. This is easy to see, but a lot harder to prove. In any case, problems of this type allow us to use a golden-section search algorithm to find this maximum. This is an iterative algorithm which finds the value which maximises a function. In this case, we choose the function we want to maximise as
\begin{equation}
    M(s) = s*E(s) \label{eqn:max_dust}\\
\end{equation}
where
\begin{equation}
    E(s) = \begin{cases}
        1 \ \ \mathrm{if} \ s \leq s_{\textrm{entr}}\\
        0 \ \ \mathrm{if} \ s > s_{\textrm{entr}}
    \end{cases}
\end{equation}
and $s_{\rm entr}$ is defined as the maximum grain radius that is entrained in the wind. Practically in the code, this entrained size is found by testing whether the particle is or isn't at the escape velocity at the end of its simulation time. The function $M(s)$ is maximised by the maximum grain size that can be entrained in a wind (function shown in Figure \ref{fig:dustsolver}).

The golden-section search method starts by setting the lower ($s_{\textrm{min}}$) and upper bounds ($s_{\textrm{max}}$), effectively starting with any size that is entrained for the lower bound, and any size that isn't for the upper bound. A set of intermediate bounds are then calculated using
\begin{align}
    s_- &= s_{\rm  max} - \frac{s_{\rm max}-s_{\rm min}}{\phi}\\
    s_+ &= s_{\rm min} + \frac{s_{\rm max}-s_{\rm min}}{\phi}
\end{align}
where $\phi$ is the golden ratio (not the azimuthal coordinate).

The algorithm can be transcribed to code as follows:
\begin{verbatim}
    if M(s_-) >= M(s_+):
        s_max = s_+
    else:
        s_min = s_-
\end{verbatim}
This is run until $s_\mathrm{min}$ and $s_\mathrm{max}$ are within the desired precision. For the rest of this paper, we solve to within 0.1\% relative error between the two values, and take the average of them as the final value.

The bound selection can be optimised by exploiting that un-entrained dust returns a value of zero, potentially short-circuiting one of the computations: 
\begin{verbatim}
    if M(s_-) == 0:
        s_max = s_-
    else:
        if M(s_+) != 0:
            s_min = s_+
        else:
            s_min = s_-
            s_max = s_+
\end{verbatim}
A final minor optimisation can be made by taking into account that larger dust sizes tend to have larger timestep sizes, and therefore take fewer steps to reach the end of simulation time. We take advantage of this by reordering the if statement to calculate the entrainment of the larger dust size first:
\begin{verbatim}
    if M(s_+) != 0:
        s_min = s_+
    else:
        if M(s_-) == 0:
            s_max = s_-
        else:
            s_min = s_-
            s_max = s_+
\end{verbatim}
With an algorithm to determine the maximum entrained dust size, we can now benchmark entrainment in a deterministic manner.




\appendix


\bsp	
\label{lastpage}
\end{document}